\documentclass[twocolumn, twocolappendix]{aastex631}

\usepackage{amsmath}
\usepackage{comment}

\shorttitle{Lensed young star cluster Transients in A370}
\shortauthors{Li et al.}

\graphicspath{{./}{figures/}}

\begin{document}

\title{Flashlights: Transients among Gravitationally-Lensed Star Clusters in the Dragon Arc.  \\ I. Stellar Microlensing vs Stellar Outbursts}

\author[0000-0002-4490-7304]{Sung Kei Li}
\email{keihk98@connect.hku.hk}
\affiliation{Department of Physics, The University of Hong Kong, Pokfulam Road, Hong Kong}

\author[0000-0001-9065-3926]{Jose M. Diego}
\affiliation{IFCA, Instituto de F\'isica de Cantabria (UC-CSIC), Av. de Los Castros s/n, 39005 Santander, Spain}

\author[0000-0003-3142-997X]{Patrick L. Kelly}
\affiliation{Minnesota Institute for Astrophysics, 
University of Minnesota, 116 Church St. SE, Minneapolis, MN 55455, USA}

\author[0000-0003-4220-2404]{Jeremy Lim}
\affiliation{Department of Physics, The University of Hong Kong, Pokfulam Road, Hong Kong}

\author[0000-0003-1060-0723]{WenLei Chen}
\affiliation{Department of Physics, Oklahoma State University, 145 Physical Sciences Bldg, Stillwater, OK 74078, USA}

\author[0000-0003-1276-1248]{Amruth Alfred}
\affiliation{Department of Physics, The University of Hong Kong, Pokfulam Road, Hong Kong}

\author[0000-0002-6039-8706]{Liliya L.R. Williams}
\affiliation{Minnesota Institute for Astrophysics, 
University of Minnesota, 116 Church St. SE, Minneapolis, MN 55455, USA}

\author[0000-0002-8785-8979]{Thomas J. Broadhurst}
\affiliation{Department of Theoretical Physics, University of Basque Country UPV/EHU, Bilbao, Spain}
\affiliation{Ikerbasque, Basque Foundation for Science, Bilbao, Spain}
\affiliation{Donostia International Physics Center, Paseo Manuel de Lardizabal, 4, San Sebasti\'an, 20018, Spain}

\author[0000-0002-7876-4321]{Ashish K. Meena}
\affiliation{Department of Physics, Ben-Gurion University of the Negev, PO Box 653, Be’er-Sheva 8410501, Israel}

\author[0000-0002-0350-4488]{Adi Zitrin}
\affiliation{Department of Physics, Ben-Gurion University of the Negev, PO Box 653, Be’er-Sheva 8410501, Israel}

\author[0000-0003-3878-9118]{Alex Chow}
\affiliation{Department of Physics, The University of Hong Kong, Pokfulam Road, Hong Kong}

\author[0000-0003-3460-0103]{Alexei V. Filippenko}
\affiliation{Department of Astronomy, University of California, Berkeley, CA 94720-3411, USA}


\begin{abstract}

We report the discovery of transients among star clusters in a distant galaxy that is gravitationally lensed by a foreground galaxy cluster, and explore whether these transients correspond to: (i) intrinsic variations associated with stellar outbursts; or (ii) extrinsic variations imposed through microlensing by intraclusters stars along, perhaps, with primordial black holes.  From images at two epochs separated by nearly a year, we discovered ten such transients -- displaying brightness variations of $\sim$10\%--20\% -- among 55 persistent knots identified as young star clusters in the Dragon arc. 
Two of these transients are associated with a triply-lensed star cluster, permitting a test of intrinsic variability by checking whether their light variations are repeated among the different lensed counterparts with a suitable time delay given their different light arrival times at the observer.  Despite considerable care in constructing a lens model for Abell 370 that is optimized at the Dragon arc, we found that the predicted lensing magnifications are not sufficiently accurate to provide a definitive test of intrinsic variability based only on two images -- although such a test will become feasible as more observations are made.  On the other hand, we perform simulations demonstrating that the observed level of brightness variations, as well as the observed transient event rate, can be explained entirely by stellar microlensing: as stars in the background star cluster move across the sky relative to intracluster stars, changes in their individual brightnesses can result in an overall change in the brightness of their parent star cluster. 
\end{abstract}


\section{Introduction} \label{sec:intro}

First mentioned by \citet{Hoag_1981} and \citet{Lynds_1986} with the first detailed published account by \citet{Soucail_1987a}, the Dragon arc holds the distinction of being the first giant arc discovered towards a galaxy cluster (Abell 370).  It also is among the first such arcs eventually recognized to be generated through gravitational lensing of a background galaxy by a galaxy cluster \citep[e.g., see brief historical description in][]{Bergmann_1990}.  Galaxy clusters are now regularly employed as cosmic lenses to search for galaxies in the distant and early Universe, and to enable measurements of their sizes if not studies of their morphologies.  For these purposes, the deepest and sharpest images of cluster cosmic lenses are provided by the {\it Hubble Space Telescope (HST)}, in particular through the Hubble Frontiers Fields (HFF) program \citep{Lotz_2017}, and by the {\it James Webb Space Telescope (JWST)}.  Indeed, the first scientific image released by the {\it JWST} is that of the galaxy cluster SMACS J0723.3–7327, highlighting a bounty of arcs corresponding to multiply-lensed images of numerous background galaxies \citep[see][]{Chow2024}.  This image affirmed the promise of the JWST for revealing galaxies formed soon after the Big Bang, whose light at ultraviolet to optical wavelengths is redshifted all the way into the infrared.

Just 6 years ago, a new class of gravitationally-lensed phenomenon was discovered towards galaxy clusters: unresolved features in lensed arcs detected sporadically as short-lived brightenings, henceforth referred to as (lensed) transients.  The first such transient was discovered by \citet{Icarus} in multi-epoch images of the cluster MACS\,J1149+2223, taken by the {\it HST} for the purpose of studying SN\,Refsdal (the first multiply-lensed supernova discovered).  Dubbed ``Icarus", this transient was located near a critical curve -- at which the lensing magnification reaches a maximum -- as predicted by lens models for MACS\,J1149+2223.  Based on its spectral energy distribution (SED), Icarus was attributed to the gravitational lensing of a single blue supergiant star at $z = 1.49$.  The lensing magnification imposed on this star reached a factor of over 2000 (placing an upper limit on its size of $\lesssim 0.06 \, \rm pc$), emphasizing the extreme lensing magnifications provided by galaxy clusters.

Having virial masses of up to $10^{15} \, M_{\odot}$, the maximal lensing magnifications provided by galaxy clusters are inversely proportional to the square root of the size of the lensed object, $\mu_{\rm max} \propto 1/\sqrt{R_{\rm obj}}$.  For background stars located at caustics so that their images form at critical curves, $\mu_{\rm max}$ can reach -- in principle -- factors of order $10^6$ \citep{Miralda_1991}.  In practice, the ubiquitous presence of microlenses (such as intracluster stars or hypothetical primordial black holes) or millilenses (e.g., intracluster globular clusters or relatively low-mass Dark Matter halos) lowers $\mu_{\rm max}$ to $\lesssim 10^5$ \citep{Venumadhav_2017,Diego_2018}.  The sudden brightening of Icarus was attributed to a change in its lensing magnification as the background blue supergiant star moved across the sky relative to the foreground intracluster stars: as this background star crossed a caustic associated with a stellar microlens and the cluster macrolens combined, it brightened dramatically before fading to obscurity.  We henceforth refer to such events as caustic-crossing events.

Not long after the discovery of Icarus, more transients attributed to casutic-crossing events were discovered, including ``Spock" \citep{Spock} and ``Warhol" \citep{Chen_2019}.  These discoveries have inspired us to initiate a dedicated and systematic study of lensed transients towards the most powerful cosmic lenses known.  Involving observations with the {\it HST}, our program -- which we refer to as Flashlights \citep{Kelly_2023a} -- targets the six galaxy clusters observed in the HFF program in two visits separated by approximately one year.  During each epoch, images of the individual cluster are acquired with the two widest filters, UVIS F200LP and F350LP, available on the WFC3.  For a given cluster, all the images are taken at the same telescope roll angle to avoid, when differencing images taken at different epochs, localized changes in brightness owing to the rotation of the telescope point-spread-function (PSF).  Such artificial changes in brightness can easily be mistaken for transients.

One of the scientific objectives of Flashlights is to collect better statistics for caustic-crossing events.  The rate of such events can provide greater insights into the nature of distant lensed stars, as well as the physical properties of the intervening microlenses and millilenses in galaxy clusters.  For example, the caustic-crossing event rate can be used to infer the abundance of luminous stars hosted by galaxies in the early universe \citep[e.g.,][]{Schauer_2022, diego2023buffaloflashlights}.   The caustic-crossing event rate also can be used to probe the properties of intracluster stars, and perhaps even the very nature of dark matter (DM).   For example, \citet{Icarus} simulated light curves for caustic-crossing events based on different assumptions for the properties of intracluster stars, as well as for a hypothetical population of primordial black holes (PBHs; a DM candidate).  They find that intracluster stars having a \citet{Salpeter} mass function, where the abundance of low-mass stars follows a power-law distribution, is preferred over the \citet{Chabrier} mass function, where the abundance of low-mass stars follows a log-normal distribution, for reproducing the observed light curve of Icarus.  Based also on the light curve of Icarus, \citet{Diego_2018} and \citet{Oguri_2018} placed upper limits on the fractional mass in DM that might comprise PBHs having masses over the range $\sim$1--$100 \, M_{\odot}$.  \citet{Diego_2018} found that there would be significantly more peaks in the light curve of Icarus than is observed if PBHs constitute more than 3\% by mass of DM in MACS\,J1149+2223.  By comparison, \citet{Oguri_2018} found that the lensing magnification would be saturated at the position of Icarus, such that it would never have had sufficient magnification to be detectable, if PBHs constitute more than 10\% by mass of DM in the cluster.

\begin{figure*}[ht!]
    \centering
    \includegraphics[width = \textwidth]{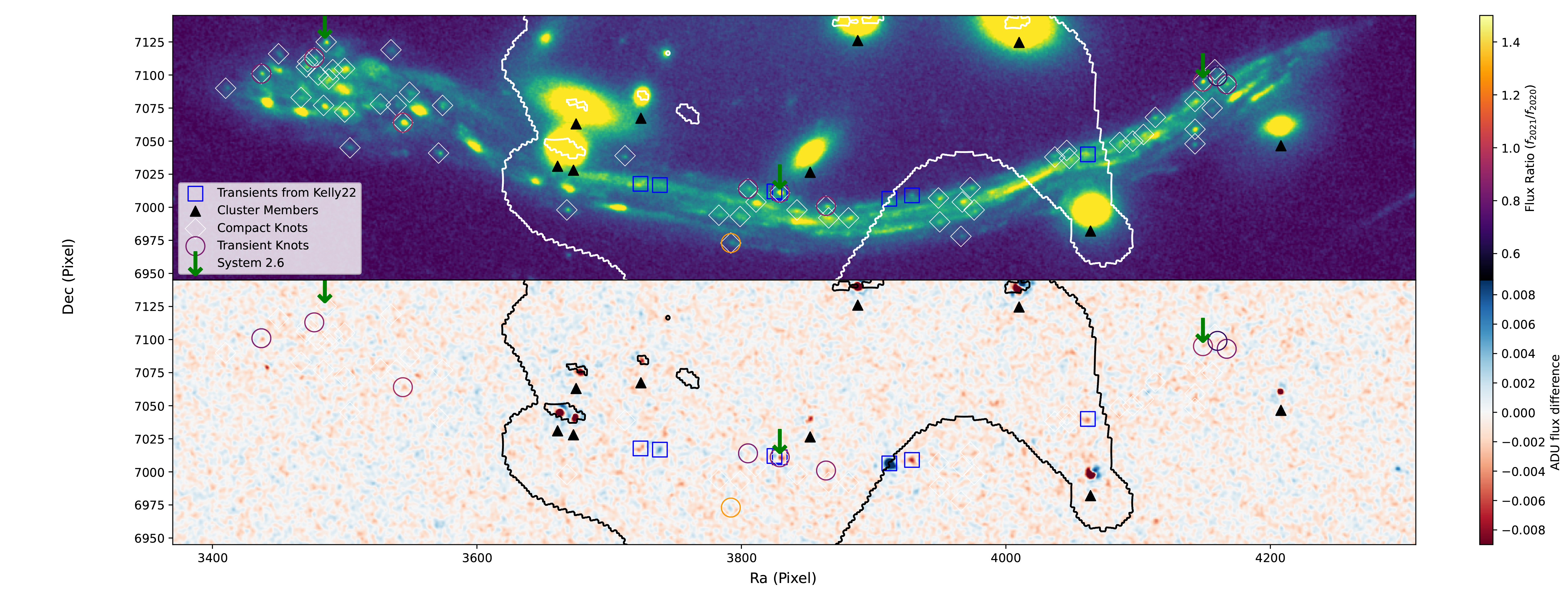}
    \caption{Upper panel: Flashlights F200LP image of the Dragon arc in 2021.  Lower panel: After subtracting the corresponding Flashlights 200LP image taken nearly a year earlier in 2020.  In both panels, the critical curve predicted by our lens model at the redshift of the multiply-lensed galaxy that forms the Dragon arc is shown in white in the upper panel and black in the lower panel.  Blue squares enclose transients reported by \citet{flashlights} that are attributed to caustic-crossing events associated with a single lensed star in the Dragon arc (see text).  In the upper panel, white diamonds enclose persistent knots that we identify as young star clusters in the Dragon arc (see text).  Knots enclosed by color circles in both panels display brightness variations $\ge 3\sigma$ between the two epochs; the color of the circle indicates the ratio in brightness of the enclosed knot between the two epochs, as encoded by the color bar to the right that spans both panels.  Two of these transients coincide with a triply-lensed star cluster (System 2.6), for which the three lensed counterparts are indicated by green arrows.  Black triangles point vertically at noise peaks associated with the bright centres of cluster members, at which the Poisson noise is significantly higher than in the rest of the image.}
    \label{fig: transient_detection}
\end{figure*}

In this paper, we focus on Flashlights observations of the galaxy cluster Abell 370 (A370), and in particular on the Dragon arc.  The two visits of this cluster in the Flashlights program are separated by 331 days (at MJD 58907 and 59236), for which the earlier image was taken about 4\,yr after the majority of images for this cluster were taken in the HFF program (MJD $\approx 57400$).  Multiple transients have been detected based on a comparison of the two Flashlights images alone \citep{flashlights, Meena_2023_offcaustic}, almost all of which are located in the Dragon arc.  The latter has been shown to correspond to a late-type galaxy at a redshift of $z = 0.7251$ \citep{Soucail_1988, GLASS, Lagattuta_2017, Chema_A370_2018}, compared to a redshift for A370 of $z = 0.375$ \citep{Struble_1999}.  As expected, many of the transients in the Dragon arc are located close to the cluster critical curves as predicted by various lens models for A370.  Furthermore, there are no detections of compact features at their positions during the HFF epoch, making them likely candidates for caustic-crossing events.

Rather than the candidate caustic-crossing events described above, in this paper we focus on transients coincident with bright and persistent knots in the Dragon arc.  These knots are all spatially unresolved, and located farther away from the cluster critical curves than the candidate caustic-crossing events -- their brightness variations therefore represent a different type of lensing transient than the caustic-crossing events described above.  We show that all the knots associated with these transients, along with other comparably bright knots having similar colours but which do not display detectable light variations, almost certainly correspond to star clusters that are still sufficiently young to display H\,II regions.  We classify those that display a change in brightness at levels $\ge 3\sigma$, where $\sigma$ is the measurement uncertainty of their individual brightness, as transients.  In all, we find a total of ten such transients, including the single transient previously reported by \citet[][]{flashlights}, in the Dragon arc.  In the upper panel of Figure~\ref{fig: transient_detection}, showing the difference between the F200LP images taken in the two Flashlights epochs, these transients are enclosed by circles; the colour of the individual circles indicates the level of brightness change associated with the enclosed knot.  The positions, changes in apparent magnitude, and lensing magnifications (based on our lens model for A370 as described later) of the knots associate with transients are compiled in Table~\ref{tab: transient_detection}.  In the lower panel of Figure~\ref{fig: transient_detection}, knots that we identified as young star clusters in the Dragon arc are enclosed by diamond symbols.  Squares enclose caustic-crossing events as identified by \citet{flashlights}.

%

\begin{deluxetable}{c|c|c|c|c}
\centering
\tabletypesize{\scriptsize}
\tablewidth{0pt} 
\tablecolumns{6}
\tablecaption{Transients among Young Star Clusters}
\tablehead{\colhead{No.}& \colhead{$\Delta \alpha$}   & \colhead{$\Delta \delta$} & \colhead{$\Delta m$}  & \colhead{$|\mu|$}}
\startdata
1 & 10\farcs88 & -6\farcs67 & 0.19 $\pm$ 0.04 & 7.90\\
2 & 9\farcs68 & -6\farcs30 & 0.13 $\pm$ 0.04  & 9.51\\
3 & 7\farcs67 & -7\farcs78 & 0.07 $\pm$ 0.02 & 14.45\\
4 & 0\farcs23 & -10\farcs51 & -0.26 $\pm$ 0.04  & 27.97\\
5 & -0\farcs16 & -9\farcs27 & 0.18 $\pm$ 0.05  & 40.01\\
{6} & -0\farcs84 & -9\farcs38 & 0.16 $\pm$ 0.02  & 13.70\\
7 & -1\farcs93 & -9\farcs67 & 0.14 $\pm$ 0.04  & 18.63\\
{8} & -10\farcs43 & -6\farcs86 & 0.09 $\pm$ 0.02 & 10.40\\
9 & -10\farcs82 & -6\farcs73 & 0.44 $\pm$ 0.08  & 9.77\\
10 & -11\farcs00 & -6\farcs92 & 0.19 $\pm$ 0.06  & 9.75\\
\enddata
\label{tab: transient_detection}
\tablecomments{$\Delta \alpha$ and $\Delta \delta$ relative to ($\alpha = 39.97134$, $\delta = -1.5822597$).  No. 6 and 8 coincide with two of the lensed counterparts of triply-lensed star cluster.}
\end{deluxetable}

What is the nature of transients associated with the lensed star clusters?  In this paper, we investigate two possibilities, for which they first is whether they are produced by energetic outbursts associated with a single luminous star in a star cluster.  As two of these transients coincide with different image counterparts of the same multiply-lensed star cluster, we test whether a transient seen in one of the image counterparts is repeated in another image counterpart after taking into consideration the different light arrival times from different image counterparts.  The second possibility we investigate is whether these transients are produced by the collective effects of a population of microlenses -- whether it be intracluster stars acting alone or in concert with PBHs -- on the population of background stars that constitute a star cluster; as the background stars moves across the sky relative to the foreground microlenses, the lensing magnification of individual stars change such as to produce a collective change in their total brightness.  The effects of such collective microlensing on a star cluster is not well understood.  \citet{Dai_2021} showed that star clusters having masses of $\sim$$10^{6-7} \,M_{\odot}$ could exhibit a very small level of flux variation ($\sim$1\%) owing to the effects of collective microlensing.  By comparison, the estimated masses of the young star clusters under study here is 2--3 orders of magnitude lower.  The effect of collective microlensing may be even more pronounced in lower-mass star clusters owing to the fewer number of luminous stars in such star clusters (provided they have comparable ages): in this situation, one or just a few luminous stars can dominate the light from a star cluster, such that even relatively small changes in their lensing magnifications can appreciably alter the total brightness of a star cluster.

The remainder of this paper is organized as follows.  In Section~\ref{sec: data}, we describe the data used in our work, for which images taken by the Flashlights program are supplemented by others taken also by the {\it HST} in two other programs.  In addition, we also used imaging spectra taken by the Very Large Telescope ({\it VLT}).  Then, in Section~\ref{sec: lens_model}, we describe how we constructed a lens model for A370 that is optimized over the Dragon arc.  In Section~\ref{sec: Intrinsic}, we use this lens model to explore whether two of the transients associated with a triply-lensed star cluster could be generated by a stellar outburst.  After that, in Section~\ref{sec: Extrinsic}, we perform simulations to test whether the observed level of brightness variations, as well as the observed transient event rate, can be generated through microlensing by intracluster stars in concert with, perhaps, primordial black holes: as stars in the background star cluster move across the sky relative to the microlenses, changes in their individual brightnesses can result in an overall change in the brightness of their parent star cluster.  In Section~\ref{sec: Discussion}, we briefly discuss how the results of our simulations might change in the presence of yet another type of intracluster object: globular clusters.  Finally, we present a concise summary of our work in Section~\ref{sec: conclusions}.  Throughout this paper, we adopt magnitudes in the AB system \citep{AB_Mag}, along with the standard cosmological parameters $\Omega_{m} = 0.3$, $\Omega_{\Lambda} = 0.7$, and H$_{0} = 70\ \textrm{km s}^{-1} \textrm{Mpc}^{-1}$.

\section{Data} \label{sec: data}

\subsection{Flashlights}

As mentioned in Section\,\ref{sec:intro}, the Flashlights program targeted the six galaxy clusters observed in the HFF program by acquiring even deeper images of these clusters with {\it HST} using the WFC3 in two filters, UVIS F200LP and UVIS F350LP, separated by $\sim 1$\,yr.  Owing to a change in the initially proposed filters for the observations (from the ACS/WFC Clear to WFC3 UVIS F350LP), images of A370 were acquired only in the UVIS F200LP filter, taken on 2020 Feb. 27 and 2021 Jan. 22 (UTC dates are used throughout this paper).  
Each image has an effective exposure time of 6.07\,hr and reached a $5\sigma$ detection threshold of $m_{\rm AB} = 30$\,mag in individual pixels with a pixel size of 30\,mas on a side. The F200LP filter spans the wavelength range 1990--10,550\,\AA, spanning observed wavelengths from the ultraviolet (UV) to the near-infrared (NIR), as we do not know a priori what color transients may have. The Flashlights images cover the same field of view (FOV) as the HFF observations of $\sim $7 arcmin$^{2}$.

\subsection{Hubble Frontier Fields}

Next to the Flashlights program, the deepest {\it HST} images of A370 were obtained by the HFF program \citep{Lotz_2017}, where images were taken with the ACS camera in the F435W, F606W, and F814W filters, as well as with the WFC3 camera in the F105W, F125W, F140W, and F160W filters.  Note that there is therefore no overlap in filter coverage between the HFF and Flashlights programs. The HFF images of A370 were taken between December 2015 and February 2016.  Just like in the Flashlights program, A370 was observed at a common telescope roll angle in all the images taken for the HFF program.  We used the final stacked images retrieved from the HFF archive\ (\url{https://archive.stsci.edu/prepds/frontier/}) for much of our work, although we also made images taken at individual epochs to measure the light curves described in Section\,\ref{sec: Intrinsic}.  These images have a pixel size of 30\,mas, the same as in the Flashlights program.

\subsection{Other HST Observations}
\label{sec: Data_HST}
{\it HST} images of A370 also were acquired as part of the Beyond Ultra-deep Frontier Fields And Legacy Observations (BUFFALO; GO-15117) program.  The BUFFALO observations of A370 comprise four exposures taken with both the F606W ($\sim 20$\,min per image) and F814W ($\sim 10$\,min per image) filters between 2018 Dec. 19 and 2019 Jan. 30, the last observation made just over a year before the beginning of the Flashlights observations of the same cluster (albeit in different filters).  The FOV of the BUFFALO images of A370 is $\sim 36\,\rm arcmin^{2}$, over five times larger than that of the HFF and Flashlights images. To produce images having the same pixel size as the Flashlights and HFF images, we recalibrated the BUFFALO images, realigned them with TweakReg, and then resampled the images to a pixel size of 30\,mas using AstroDrizzle.  The BUFFALO images were used together with the HFF and Flashlights images to assemble the light curves described in Section\,\ref{sec: Intrinsic}.

\subsection{MUSE Spectra} \label{sec: catalog}

Correct identifications of multiply-lensed counterparts along with accurate determinations of their redshifts are crucial starting points for constructing reliable lens models.  Furthermore, the identification of cluster members is necessary to provide one of the ingredients for lens models of galaxy clusters.  For these purposes, we used spectroscopic measurements of A370 from observations with MUSE on the VLT. MUSE is an integral field unit (IFU) with a FOV of 1\,arcmin$^{2}$, providing images with a pixel size of 0.2\arcsec \ over the wavelength range 4650--9300\,\AA.

The latest release of the MUSE data for A370 covers an FOV of $\sim 4$\,arcmin$^{2}$ that is comparable to the estimated angular size of A370 \citep{Lagattuta_2019}.  Although smaller than the FOV of the aforementioned {\it HST} observations of A370, it is nevertheless sufficiently large to span the Einstein ring of the cluster and beyond, including the Dragon arc.  Notice that the pixel size of the MUSE images ($0.2\arcsec$) is much larger than that of the HFF images ($0.03\arcsec$), in keeping with the much larger PSF of the MUSE images ($\sim 0.8\arcsec$) compared with the {\it HST} images. To remove as best as possible night-sky lines, we applied ZAP \citep{ZAP} to the images after masking out cataloged sources.


\section{Lens Modeling} \label{sec: lens_model}

\begin{figure*}[t!]
    \centering
    \includegraphics[width = .9\textwidth]{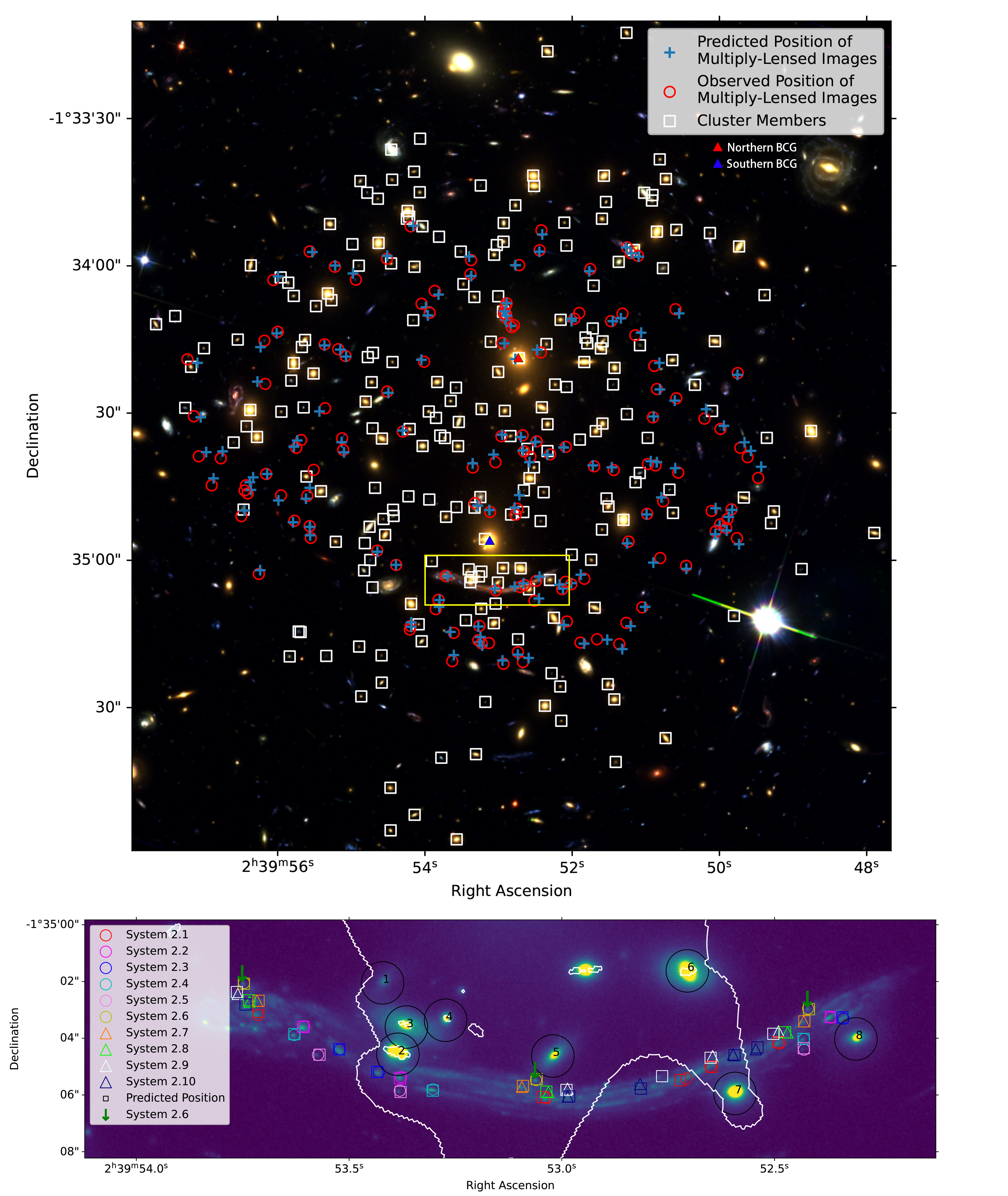}
    \caption{Upper panel -- RGB image of A370 constructed from {\it HST} images taken in the F435W (blue), F606W (green), and F814W (red) filters.  White squares enclose cluster members.  Red circles enclose multiply-lensed images of systems used to constrain our lens model for A370; blue crosses enclosed by or closely adjacent to red circles indicate positions predicted by our lens model for these images.  Yellow rectangle encloses the Dragon arc.  Lower panel: close-up of the Dragon arc.  White line traces the critical curve at the redshift of the multiply-lensed galaxy that forms the Dragon arc.  Circles having a common color enclose lensed counterparts of knots identified through their common colours and MUSE-VLT spectra, as well as from delensing and relensing.  Triangles having a common color enclose lensed counterparts of knots too dim for color or spectra analyses, and which are identified solely through delensing and relensing.  All these knots were used as constraints for optimizing our lens model at the Dragon arc.  Green arrows point to the three lensed counterparts of System 2.6, two of which are among the transients detected.  Black circles enclose cluster members that are parameterised individually for optimizing our lens model for A370 at the Dragon arc; the numbers in the circles correspond to labels for cluster members as referred to in Table\,\ref{tab: global_model}.}
    \label{fig: RGB}
\end{figure*}

The most recent lens models of A370 are those reported by \citet[][hereafter D18]{Chema_A370_2018} and \citet[][hereafter L19]{Lagattuta_2019}.  In addition to a common subset of 30 multiply-lensed systems serving as constraints on their lens models, D18 also used two sets of multiply-lensed knots that they identified in the Dragon arc, whereas L19 also used 15 other multiply-lensed systems that they identified.  In our work, we started with the same set of multiply-lensed systems identified by L19, comprising a total of 136 lensed images --- but ended up combining two systems into one, as well as omitting another, for reasons explained in Section~\ref{sec: constraints}.  Furthermore, to provide more accurate lensing predictions in the region of the Dragon arc, we optimized our lens model over this region by using as additional constraints nine sets of multiply-lensed knots in the Dragon arc. The manner by which these multiply-lensed knots were identified is described in Section~\ref{sec: constraints}.

\subsection{Ingredients}

Our lens model consists of mass components representing the individual cluster members as well as those representing the cluster-scale DM halos.  As previously found by L19, multiple cluster-scale DM halos are necessary to reproduce the multiply-lensed images generated by this dynamically disturbed cluster.  We constructed our lens model using the lens modeling algorithm \texttt{Glafic} \citep{glafic}.

To select cluster members, we compiled a list of all galaxies lying in the redshift interval $z = 0.3745 \pm 0.0231$ (i.e., $\pm 3\sigma$ from the center of a Gaussian distribution in radial velocities; L19).  Redshifts were taken from the spectroscopic redshift catalog compiled by \citet{Richard_2021}, all based on observations with MUSE.  Out of a total of 222 galaxies that satisfied the selection criteria, we omitted 4  flagged as having poor photometric quality, thus leaving 218 spectroscopically-identified cluster members.  In the lens modeling, we parameterized all the selected cluster members, apart from the two brightest cluster galaxies (BCGs), using a modified Jaffe profile as described by \citet{Keeton_2001}, with a truncation radius and mass scaled to their individual luminosities \citep[following the procedure described by][]{Chow2024}.  In the preliminary lens model, we forced all these cluster members to share the same mass-to-light ratio ($M/L$); when optimizing our lens model over the Dragon arc, we allowed the eight cluster members in close vicinity (those enclosed by black circles in the bottom panel of Figure~\ref{fig: RGB}) to have varying $M/L$ ratios. To parameterize the two BCGs, we adopted Sersic profiles \citep{Sersic}, for which their $M/L$ and Sersic indices were permitted to be independent of each other and to be freely solved.  We allowed these freedoms as several multiply-lensed images lie in their close vicinity, thus placing more stringent constraints on their mass profiles. The centroid positions, ellipticities, and position angles of all these cluster members were taken from \citet{Richard_2021}. 

Numerous studies suggest that A370 is likely to be in the midst of a merger between two (or more) galaxy clusters \citep[e.g.,][]{Oemler_1997, De_Filippis_2005, Richard_2010, Strait_2018, Molnar_2020, Ghosh_2021, de_Oliveira_2022}.  Thus motivated, we began by implementing into our lens model two cluster-scale DM  haloes having Navarro-Frenk-White (NFW) profiles \citep{Navarro_1997, Oguri_anfw}. In so doing, we found relatively large discrepancies between the predicted and observed positions among many of the multiply-lensed systems used as constraints. Adding one more DM halo, also having an NFW profile, significantly improved upon the agreement between the predicted and observed positions of the multiply-lensed systems. Adding yet another DM halo provided no improvement in agreement. Our lens model therefore comprises three DM halos, all having NFW profiles, for A370.  For comparison, L19 incorporated four DM halos into their lens model.

To allow for the possibility of gravitational perturbations produced by an inhomogeneous distribution of matter around the cluster, we also explored whether including a shear term \citep{Keeton_1997} in our lens model improved the agreement between the predicted and observed positions of the multiply-lensed systems used as constraints.  In the lens modeling, we allowed the strength of the shear ($\gamma$, the stretching power), convergence ($\kappa$), and shear angle to be freely solved.  We found the best-fit lens model to require a shear term; the parameters of this model are listed in Table~\ref{tab: global_model}.  The critical curve predicted by our lens model in the region of the Dragon arc, for which our lens model is especially tuned, is shown by the white line in the bottom panel of Figure~\ref{fig: RGB}.

\begin{deluxetable*}{lcccclcc}
\label{tab: global_model}
\centering
\tabletypesize{\scriptsize}
\tablewidth{0pt} 
\tablecaption{Lens Model Parameters}
\tablehead{Component (Profile)& $\Delta \alpha$       & $\Delta \delta$      & e & $\theta$ & Mass ($M_{\odot}h^{-1}$)& c (NFW)/ n (Sersic) & $r_{\textrm{trun}}$ }
\startdata
DM1 (NFW)           & -0\farcs01   &  5\farcs40   & 0.20  & $-174^\circ.73$   &  $1.06\times 10^{15}$     & 5.11  &  --   \\
DM2 (NFW)            & 27\farcs69 & 40\farcs22     & 0.64  &  $ -7^\circ.11$ &  $9.01\times 10^{14}$     & 0.28  & --\\
DM3 (NFW)           & -5\farcs25  & 36\farcs58    & 0.32  &  $-160^\circ.29$  & $1.24\times 10^{14}$     & 7.68  & --   \\
NBCG (Sersic)           &   -5\farcs90       & 37\farcs24    &  0.20 &  $357^\circ.31$  &  $4.05\times10^{11}$    & 4.34  & 6\farcs36    \\
SBCG (Sersic)         &   0\farcs01       &  0\farcs02   & 0.30  & $315^\circ.12$   &  $3.64\times10^{11}$    & 5.18  & 3\farcs55 \\
Cluster Members &   --       &   --   &   --  &   --  &  $\sigma_{*} = 258.32 $    &  --   &     $r_{\textrm{trun},*} = 34\farcs10$\\
Shear           &  --    &   --   &  --  & $-43^\circ.40$  &   --  & $\gamma$ = 3.47e-2   & --  \\
\hline
Cluster Member 1        & -4\farcs44        & -5\farcs73   & 0.59 & $-51^\circ.93$  & $1.20 \times 10^{10}$ &  --   & 1\farcs0    \\
Cluster Member 2        & -3\farcs99       & -8\farcs35   & 0.30 & $68^\circ.17$ & $1.75 \times 10^{11}$  &  --   & 2\farcs0       \\
Cluster Member 3        & -3\farcs74        & -7\farcs35   & 0.72  & $69^\circ.52$ & $1.28 \times 10^{11}$ &  --   & 5\farcs6       \\
Cluster Member 4        & -2\farcs23        & -7\farcs17   & 0.12  & $-14^\circ.29$ & $2.00 \times 10^{11}$ &  --   & 13\farcs7       \\
Cluster Member 5        & 1\farcs62        & -8\farcs43   & 0.48  & $-49^\circ.95$ & $8.34 \times 10^{10}$ &  --   & 2\farcs0       \\
Cluster Member 6        & 6\farcs26        & -5\farcs51   & 0.34  & $71^\circ.95$ & $1.41 \times 10^{12}$ &  --   & 21\farcs1      \\
Cluster Member 7        & 7\farcs95        & -9\farcs79   & 0.13  & $75^\circ.52$ & $9.65 \times 10^{11}$ &  --   & 18\farcs3     \\
Cluster Member 8        & 12\farcs24       & -7\farcs89   & 0.23  & $-58^\circ.59$ & $1.21 \times 10^{11}$ &  --   & 9\farcs1        \\
\enddata
\tablecomments{$\Delta \alpha$ and $\Delta \delta$ relative to ($\alpha = 39.97134$, $\delta = -1.5822597$).  $\theta$ measured anti-clockwise from  North.  $\sigma_{*}$ denotes the velocity dispersion and $r_{\textrm{trun},*}$ the radius of the brightest non-BCG cluster member, scaled to their individual brightnesses.}
\end{deluxetable*}

\subsection{Constraints}
\label{sec: constraints}

To constrain our lens model, we employed nearly all the multiply-lensed images identified by L19, except for the following minor modifications.

\begin{enumerate}
    
    \item We combined systems 7 and 10. Both D18 and L19 suspect, based on their lens models, that these two systems may actually be part of a single system. Our lens model affirms systems 7 and 10 to be lensed counterparts,  consistent with their common redshift $z = 2.7512$ as determined from their MUSE spectra \citep{Richard_2021}.
    
    \item We omitted system 39, failing to reproduce the proposed multiply-lensed counterparts of this system in any of our lens models.  MUSE spectra show only the marginal detection of one emission line, if any, among the proposed multiply-lensed counterparts.
    
\end{enumerate}

A complete list of the 43 multiply-lensed systems, which together form 133 lensed images, used to constrain our preliminary lens model (i.e., before being tuned over the Dragon arc) is compiled in Table~\ref{tab: system_list} of the Appendix. Their image counterparts are enclosed within red circles in the upper panel of Figure~\ref{fig: RGB}; as can be seen, the multiply-lensed images are quite evenly distributed throughout the cluster.  Our preliminary lens model constrained by just these systems is able to reproduce all their positions within a root-mean-square (RMS) dispersion of $0\farcs88$, comparable to the RMS positional dispersions of other models like that of L19.  As a test of the predictability of our lens model, we were able to reproduce the observed flux ratios of systems 4, 9, and 26, the only systems for which we were able to extract reliable photometry (one or more lensed counterparts of the remaining systems lie close to cluster members or other bright objects), to within $3\sigma$.

To optimize our lens model over the Dragon arc, we searched for additional multiply-lensed counterparts within the Dragon arc.  As can be seen in the lower panel of Figure~\ref{fig: RGB}, the Dragon arc contains a multitude of knots, at least some of which correspond to young star clusters as we shall explain.  We searched for multiply-lensed counterparts by looking for knots having similar colors, after which we checked their spectra from MUSE.  As an additional check, we delensed and released the knots thus selected to verify that they are indeed associated based on the predicted positions of their lensed counterparts.  In this way, we identified five sets of multiply-lensed knots having different colors.  Despite their different colors, all are relatively blue by comparison to the more diffuse arcs in which they are embedded.  All also are associated with line-emitting regions characteristic of H\,II regions as observed by MUSE; at the angular resolution of the MUSE observations, however, no single knot can be uniquely identified with a given line-emitting region.  We refer to these five sets of multiply-lensed knots as systems 2.2 to 2.6.  The spectra of the line-emitting regions in which each of these knots appears to be embedded are shown in Figure~\ref{fig: spectra_all} of the Appendix.  That each and every knot of a given system is associated with a line-emitting region argues for a true association between these knots and line-emitting regions.

Note that D19 had previously identified the same lensed counterparts for systems 2.2 and 2.3, albeit without the benefit of spectroscopy.  Their work provides reassurance that these systems have been correctly identified.  The lensed counterparts of systems 2.1--2.6 are enclosed within color circles in the lower panel of Figure~\ref{fig: RGB}.  Note that system 2.1 had previously been identified as the relatively red bulge of the galaxy that is multiply-lensed to form the Dragon arc, and which was used as one of the lensing constraints in our preliminary lens model.   Now, using systems 2.1--2.6 in addition to the lensed counterparts belonging to the other systems mentioned above, we solved for a new lens model that is now especially tuned over the Dragon arc.

We used the newly tuned lens model to search for even dimmer (and therefore less obvious) lensed counterparts in the Dragon arc.  To make this possible, we first created an unsharp-masked image of the Dragon arc, and then delensing and relensing selected knots to predict the positions of their lensed counterparts.  In this way, we identified four more sets of multiply-lensed counterparts labeled as systems 2.7--2.10, which are enclosed within triangles in the lower panel of Figure~\ref{fig: RGB}.  Unfortunately, all of these proposed lensed counterparts are too dim to display useful spectra from MUSE.  Nonetheless, all are presumably star clusters.

The positions of all the lensed counterparts in systems 2.2--2.10, as well as system 2.1 for the sake of completeness (also listed in Table\,\ref{tab: system_list} in the appendix), are compiled in Table~\ref{tab: knot_counterpart}.  All these systems, along with the other multiply-lensed systems compiled in Table\,\ref{tab: system_list}, are used to constrain our final lens model for A370 -- which is therefore optimized over the Dragon arc. This lens model is able to reproduce the positions of all the multiply-lensed knots in the Dragon arc used as constraints with an rms dispersion of $0.07 \arcsec$, which is comparable with the typical size of the PSF (FWHM $ \approx 0.08\arcsec$) of the HFF images.


\begin{deluxetable}{ccc|ccc}
\label{tab: knot_counterpart}
\centering
\tabletypesize{\scriptsize}
\tablewidth{0pt} 
\tablecaption{Multiply-lensed knots in the Dragon arc}
\tablehead{System & $\Delta \alpha$   & $\Delta \delta$ & System & $\Delta \alpha$   & $\Delta \delta$}
\startdata
    2.1.1   &  8\farcs96  & -7\farcs08  & 2.6.1   &    9\farcs45 &	-5\farcs96  \\
    2.1.2   &  -1\farcs20  &  -10\farcs01 &  {\bf2.6.2}  &    -0\farcs84 &	-9\farcs38 \\
    2.1.3   & -6\farcs14   &  -9\farcs32  &  {\bf2.6.3}   &    -10\farcs43 &	-6\farcs86\\
    2.1.4   & -6\farcs99   &   -8\farcs90 & 2.7.1   &   9\farcs00 & -6\farcs57\\
    2.1.5   &  -9\farcs41  &  -8\farcs08  & 2.7.2   &   -0\farcs34 & -9\farcs62\\
    2.2.1   &  7\farcs34  & -7\farcs49  &  2.7.3   &   -10\farcs27 &  -7\farcs30 \\
    2.2.2   &  3\farcs92  &  -9\farcs27  & 2.8.1   & 9\farcs26   & -6\farcs34 \\
    2.2.3   &  -11\farcs22  &  -7\farcs17  &  2.8.2   &  -1\farcs13  & -9\farcs80\\
    2.3.1   &  6\farcs08  &  -8\farcs30 &  2.8.3   &  -9\farcs70  & -7\farcs70\\
    2.3.2   &  4\farcs70  &  -9\farcs09 & 2.9.1   & 9\farcs68    &  -6\farcs34\\
    2.3.3   & -11\farcs67   &  -7\farcs21 &   2.9.2   &   -1\farcs95 & -9\farcs68 \\
    2.4.1   &   7\farcs68 &	-7\farcs79 &   2.9.3   &  -7\farcs06  & -8\farcs53\\
    2.4.2   &    2\farcs79	& -9\farcs71   &  2.9.4   &  -9\farcs38  & -7\farcs66\\
    2.4.3   &    -10\farcs28 & -7\farcs93  &  2.10.1   & -1\farcs95   & -9\farcs96\\
    2.5.1   &    6\farcs84 &	-8\farcs49  &   2.10.2   &  -4\farcs50  &  -9\farcs51\\
    2.5.2   &    3\farcs96 &	-9\farcs75  &   2.10.3   &  -7\farcs78  &  -8\farcs48\\
    2.5.3   &   -10\farcs28 & -8\farcs30   &  2.10.4 & -8\farcs58 & -8\farcs29\\
    \hline
\enddata
\tablecomments{$\Delta \alpha$ and $\Delta \delta$ relative to ($\alpha = 39.97134$, $\delta = -1.5822597$).  System 2.6.2 and 2.6.3 host transients.}
\end{deluxetable}

\section{Intrinsic Variability?}
\label{sec: Intrinsic}

Among the many transients detected in the Dragon arc as shown in Figure\,\ref{fig: transient_detection}, two are associated with system 2.6.  These are the only transients coinciding with two or more lensed counterparts among all the multiply-lensed systems that we identified in the Dragon arc (see Section\,\ref{sec: constraints}).  In system 2.6, both knots 2.6.2 and 2.6.3 dimmed by roughly 10\% (more precisely, $\Delta m = 0.16 \pm 0.02$\,mag for knot 2.6.2 and $\Delta m = 0.09 \pm 0.02$\,mag for knot 2.6.3) between the two epochs of the Flashlights observations separated by nearly one year --- raising the possibility that their variations in brightnesses are intrinsic to the source.

\subsection{Luminous Blue Variables} \label{subsec: LBV}
We begin by ruling out novae or supernovae for the observed dimming of knots 2.6.2 and 2.6.3.  Such an event ought to have produced a comparable dimming of knot 2.6.1 after allowing for differences in the light arrival time of the three lensed counterparts of system 2.6 as predicted by our lens model (see Section\,\ref{sec: lightcurves}). 
Next to such events, the brightest stellar outbursts are produced by LBVs ---  very massive and very luminous supergiant stars that undergo irregular stellar outbursts whose causes are not fully understood \citep[see reviews by][]{Humphreys_1994, Smith_2011, Kalari_2018}.  LBVs have been invoked to explain a number of gravitationally-lensed transients reported in the literature \citep[e.g.,][]{godzilla}.  Comprising the most massive and luminous stars known in the Galaxy, LBVs have very short lifetimes and are only found in star-forming regions.  Given the relatively blue colors of system 2.6 and its spatial coincidence with line-emitting regions, this system potentially harbors one or more LBVs.

LBVs span a wide range of luminosities, with absolute magnitudes in the $V$ band of $-8.5 \lesssim M_{V} \lesssim -11$ \citep{Weis_2020}. To scale the absolute magnitude of LBVs as measured in the rest-frame V band to that which would be measured as observed through the F200LP filter, we assume that LBVs have a blackbody spectrum with an effective temperature over the range 15,000--20,000\,K \citep{Kalari_2018} depending on the phase of their variability.  In that case, for an LBV at $z = 0.7251$, its absolute magnitude as observed through the F200LP filter is comparable to that in the rest-frame UV.
In that case, for a variation of $\Delta m_{F200LP} = 0.16\ \pm\  0.02$\,mag between the two Flashlights epochs ($\sim 190$ days in the rest frame) as measured for knot 2.6.2 to be attributed to a single LBV, this star would have had to vary intrinsically in brightness by $\Delta m_{F200LP} > 0.16$\,mag as it contributes only part of the total light from a star cluster.  For a star cluster having $M_{F200LP} = -12.5$\,mag as is the case for knot 2.6.2 given its macromagnification of $\sim 13.5$ according to our lens model, an LBV having $M_{F200LP} \approx -11$\,mag would only contribute about one-quarter of its total light.  Such an LBV would then be required to vary in brightness by $\Delta m_{F200LP} \gtrsim 0.5$\,mag to produce the dimming observed for knot 2.6.2.  Such a level of brightness variation over the timescale of the two Flashlight observations is compatible with the S~Doradus variability of LBVs, during which they vary in brightness by about $\pm 1$\,mag over timescales from years to decades \citep{Kalari_2018}. 

\subsection{Light Curves}
\label{sec: lightcurves}

If the dimming of knots 2.6.2 and 2.6.3 is intrinsic to system 2.6, the lack of a similar brightness variation in knot 2.6.1 would then require light from this knot to arrive at a much different time (either earlier or later) than knots 2.6.2 and 2.6.3.  To test this possibility, we examine differences in the arrival time of light from knots 2.6.1--2.6.3 as predicted by our lens model: if the light variation is intrinsic to the source, we should see the same variation in all the lensed counterparts subject to a time delay.  Our lens model predicts light from knot 2.6.3 to arrive $\sim 178$\,days earlier than knot 2.6.2, and light from knot 2.6.1 to have arrived even earlier: $\sim 107$\,days earlier than knot 2.6.3, and $\sim 285$\,days earlier than knot 2.6.2.  Thus, an event that occurred in system 2.6 would be seen first in knot 2.6.1 before knot 2.6.3, and thereafter followed by knot 2.6.2.  Given that the separation between the two epochs of Flashlight observations is $\sim330$\,days, knot 2.6.1 should therefore have dimmed between the two epochs of the Flashlight observations --- provided that system 2.6 had undergone a gradual dimming throughout all the epochs in which its three multiply-lensed knots were observed in the Flashlights program.  

\begin{figure}[ht!]
    \centering
    \includegraphics[width = .45\textwidth]{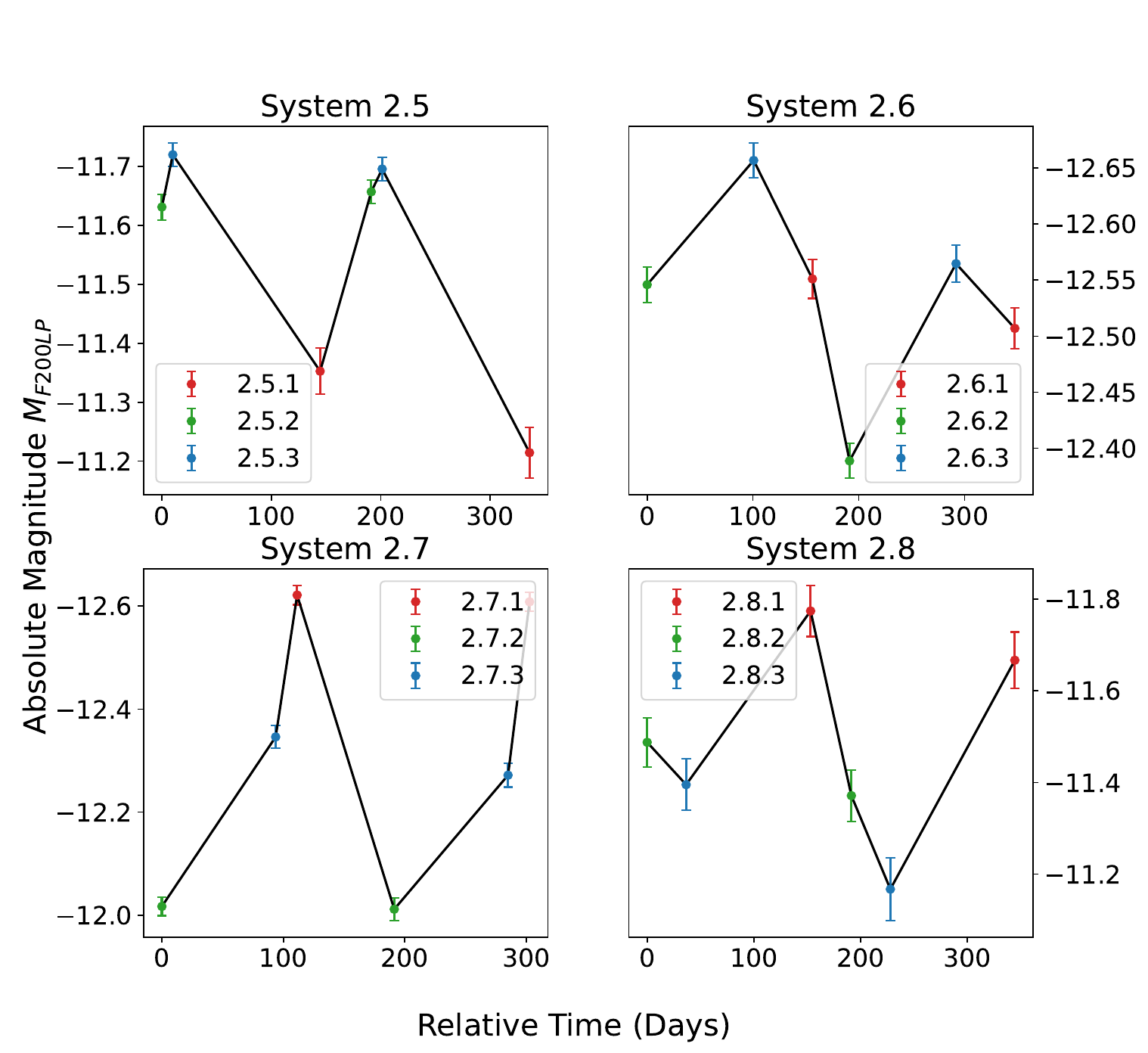}
    \caption{Light curves of Systems 2.5 to 2.8, each of which is triply lensed and for which different colors correspond to different lensed counterparts.  The brightness of each lensed counterpart has been corrected for the lensing magnification imposed by A370 as predicted by our lens model, and have different light arrival times at the observer as predicted also by our lens model.  Only system 2.6 hosts genuine transients corresponding to a change in brightness by a given lensed counterpart (the two in this system indicated in green and blue) of $\ge 3\sigma$ between the two epochs.  Yet, all the other systems display apparent light variability as judged by their light curves, owing to the limited accuracy in lensing magnifications predicted by our lens model over the Dragon arc.  As a consequence, we are not able to judge whether the genuine light variations detected among two of the lensed counterparts in System 2.6 are intrinsic to this system -- based on observations over two epochs alone, although such a test will become feasible in the future as more images are taken.  }
    \label{fig: LC_all}
\end{figure}

To test whether system 2.6 had undergone a gradual dimming, we construct a light curve for this system by applying the aforementioned time delays inferred for its individual lensed counterparts, as well as correcting for the lensing magnifications of the individual counterparts as predicted by our lens model.  The result is shown is shown in Figure~\ref{fig: LC_all}.  As can be seen, the light curve of this system displays a more complex time behavior than just a simple dimming --- such that knot 2.6.1 was, perhaps, observed at two epochs when system 2.6 just happened to have similar brightnesses.

Before drawing any further inferences from the light curve of Figure~\ref{fig: LC_all}$a$, it is prudent to construct light curves for the other knots in the Dragon arc to check on the reliability of their inferred lensing magnifications --- as even relatively small errors in the inferred lensing magnifications of different counterparts belonging to the same system can give rise to apparent time variability in their light curves constructed in the manner described above.  In Figure~\ref{fig: LC_all}, we show the light curves constructed for systems 2.5, 2.7, and 2.8, all selected for their similar lensing configurations to system 2.6.  Although none of the lensed counterparts in these systems display any significant variations in brightnesses between the two Flashlights epochs, nonetheless their light curves display a similar level of variability as system 2.6.  Their apparent time variability is therefore induced solely by corrections applied to their individual lensing magnifications as predicted by our lens model, rather than any real-time variability in the brightnesses of these systems.  Errors in the inferred lensing magnification of individual counterparts could arise from systematic errors in our lens model, or from the presence of other lensing objects not taken into account in our lens model \citep[e.g., intracluster globular clusters; see][]{Diego2024_3M}.  At the accuracy of lensing magnifications predicted by our lens model for A370, a conclusive test of intrinsic variability would require observations over multiple (much more than two) epochs spanning a longer period, so as to test whether variations in the brightness of a given lensed counterpart are reflected in the other lensed counterparts with a suitable time delay.

\begin{figure}[ht!]
    \centering
    \includegraphics[width = .45\textwidth]{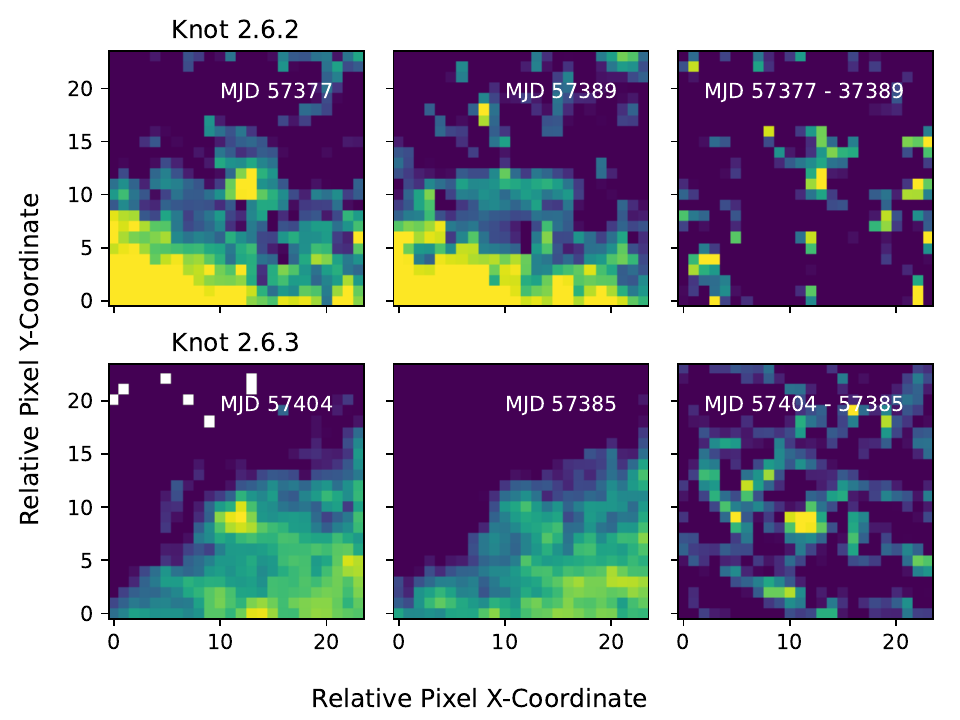}
    \caption{Brightness variations detected in knot 2.6.2 (upper row) and knot 2.6.3 (lower row) in F814W images taken during the HFF program.  In each row, first two panels are images taken at two different epochs, and last panel is the difference image between these two epochs.}
    \label{fig: HFF_example}
\end{figure}

\begin{figure*}[ht]
    \centering
    \includegraphics[width = \textwidth]{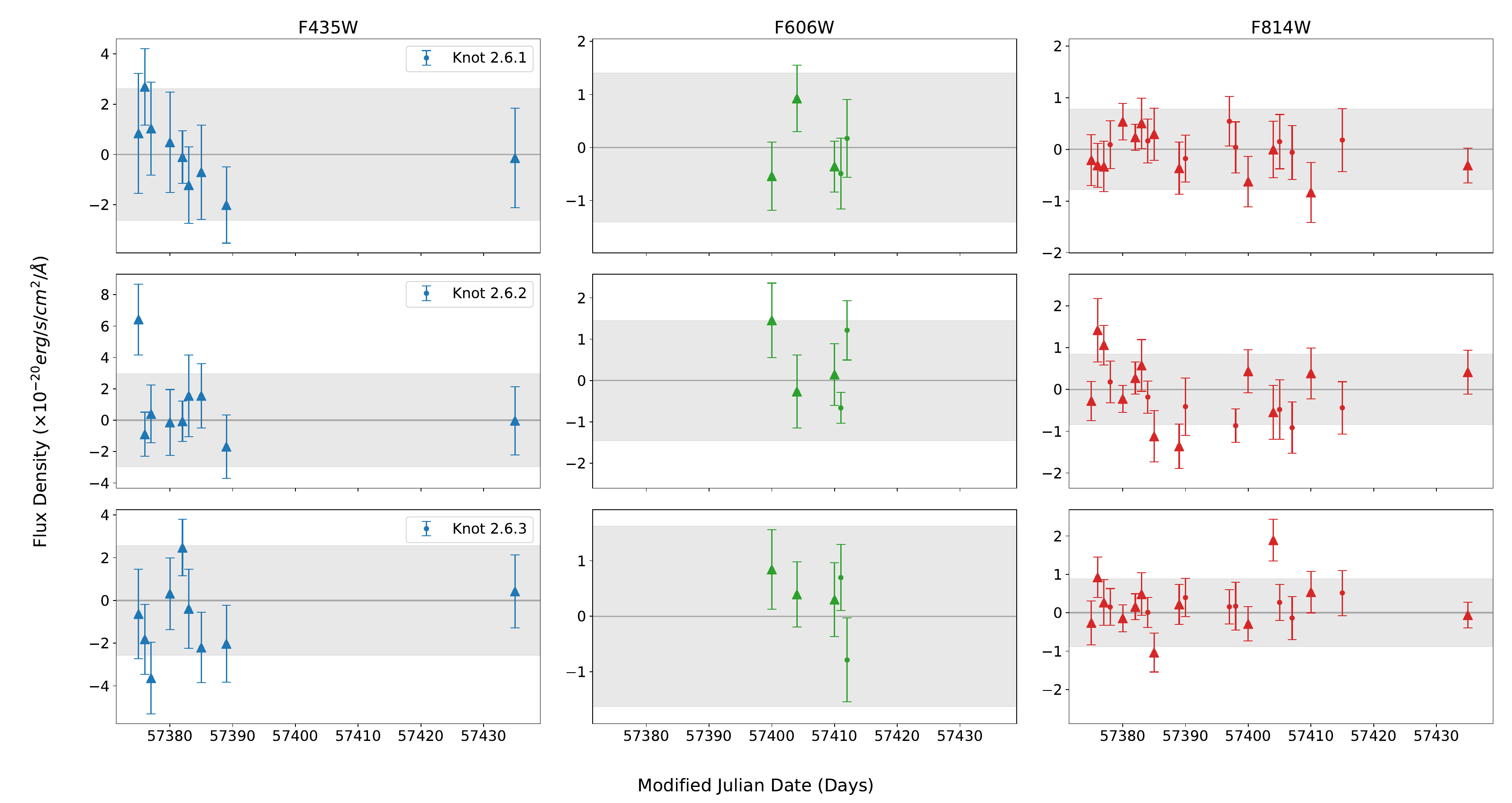}
    \caption{Light curves of the three lensed counterparts of System 2.6 in the HFF observations after subtracting their individual average brightness over the epochs shown.  Left column is brightness in the F435W filter, middle column is brightness in the F606W filter, and right column is brightness in the F814W filter.  First row is knot 2.6.1, second row knot 2.6.2, and third row knot 2.6.3.  
The gray shaded band spans $\pm 1.5\sigma$ of the average brightness.  Any observations lying beyond opposite sides of the gray band would indicate a brightness variation $> 3\sigma$.}
    \label{fig: HFF_LC}
\end{figure*}

Although the available data do not allow the aforementioned test, we can check for other examples of time variability from the HFF observations of A370, comprising observations at multiple epochs spanning about 2 months.  As mentioned in Section\,\ref{sec: data}, these observations were made about 4--5\,yr earlier than those in the Flashlights program.  Figure~\ref{fig: HFF_LC} shows the light curves for the individual knots in system 2.6 in the F435W, F606W, and F814W filters, for which observations were made with relatively even temporal cadence.  Despite the individual images being much shallow than those taken in the Flashlights program, we find instances during which the brightness of a given knot either increased or decreased by $> 3\sigma$ between one epoch and a later epoch, such as knot 2.6.2 between MJD 57377 and 57389 in F814W, and knot 2.6.3 between MJD 57385 and MJD 57404 in F814W, as confirmed by an inspection of their images and by differencing their images at the aforementioned epochs as shown in Figure~\ref{fig: HFF_example}.  As the time span of the HFF observations is much shorter than the time delay between the arrival of light from knots 2.6.2 and 2.6.3, variations in their brightnesses during the HFF observations must therefore correspond to different events (or different intervals of the same event).  Interestingly, their maximal variations in brightnesses of $\sim 0.2$\,mag are comparable to their brightness variations between the two epochs of the Flashlights observations.  Furthermore, only significant variations in brightness were observed for knots 2.6.2 and 2.6.3, whereas knot 2.6.1 showed no statistically significant variations in brightness over the entire duration of the HFF observations.

As one final test of intrinsic variability, we checked whether variations in brightness are also accompanied by variations in colour.  As lensing does not change color if involving only a single light source (e.g., one star), any changes in color might point to intrinsic variability.  We find no significant changes in color in any episodes in which we see significant variations in brightness, either in the HFF or Flashlights program, thus providing no firm evidence for (albeit not ruling out) intrinsic variability.  In the situation where transients are induced extrinsically by collective microlensing on a population of stars belonging to a background star cluster, as considered in the next section, such extrinsic variability could also be accompanied by color variations.


\section{Extrinsic Variability}
\label{sec: Extrinsic}

In this section, we explore whether the observed brightness variations of either singly- or multiply-lensed star clusters in the Dragon arc, as shown in Figure\,\ref{fig: transient_detection}, might be imposed by objects external to these star clusters.  Specifically, we examine whether their brightness variations could be induced through microlensing by relatively compact objects having stellar masses at the redshift of the galaxy cluster A370.  Such objects --- whether they be intracluster stars or PBHs --- located along the sightline to a background star cluster can alter (either increasing or decreasing) the brightness of individual stars in the star cluster, thus also altering the observed total brightness of the entire star cluster.  As the foreground lensing objects move across the sky relative to the background star clusters, the observed total brightness of these star clusters can vary with time,  thus appearing as lensed transients.

To explore the level of extrinsic variability induced by microlensing, it is more convenient to consider what happens in the source plane --- that is, at the redshift of the multiply-lensed galaxy to form the Dragon arc.  In this plane, each microlens (at the redshift of A370) produces a microcaustic; the motion of a microlens across a background star cluster is then equivalent to the motion of its microcaustic through this star cluster.  In Section~\ref{sec: simulation}, we present our microlensing simulations involving the motion of a web of microcaustics through a stellar population.  From these simulations, we derive a semianalytical approximation of the resulting variations in the total brightness of this stellar population, as described in Section~\ref{sec: analytical}.  In Section~\ref{sec: muL_explain}, we then consider whether our model predictions can explain the rate of transients observed in the Dragon arc at their observed level of brightness variations.

\subsection{Microlensing Simulation}
\label{sec: simulation}

\begin{figure*}[ht!]
    \centering
    \includegraphics[width = \textwidth]{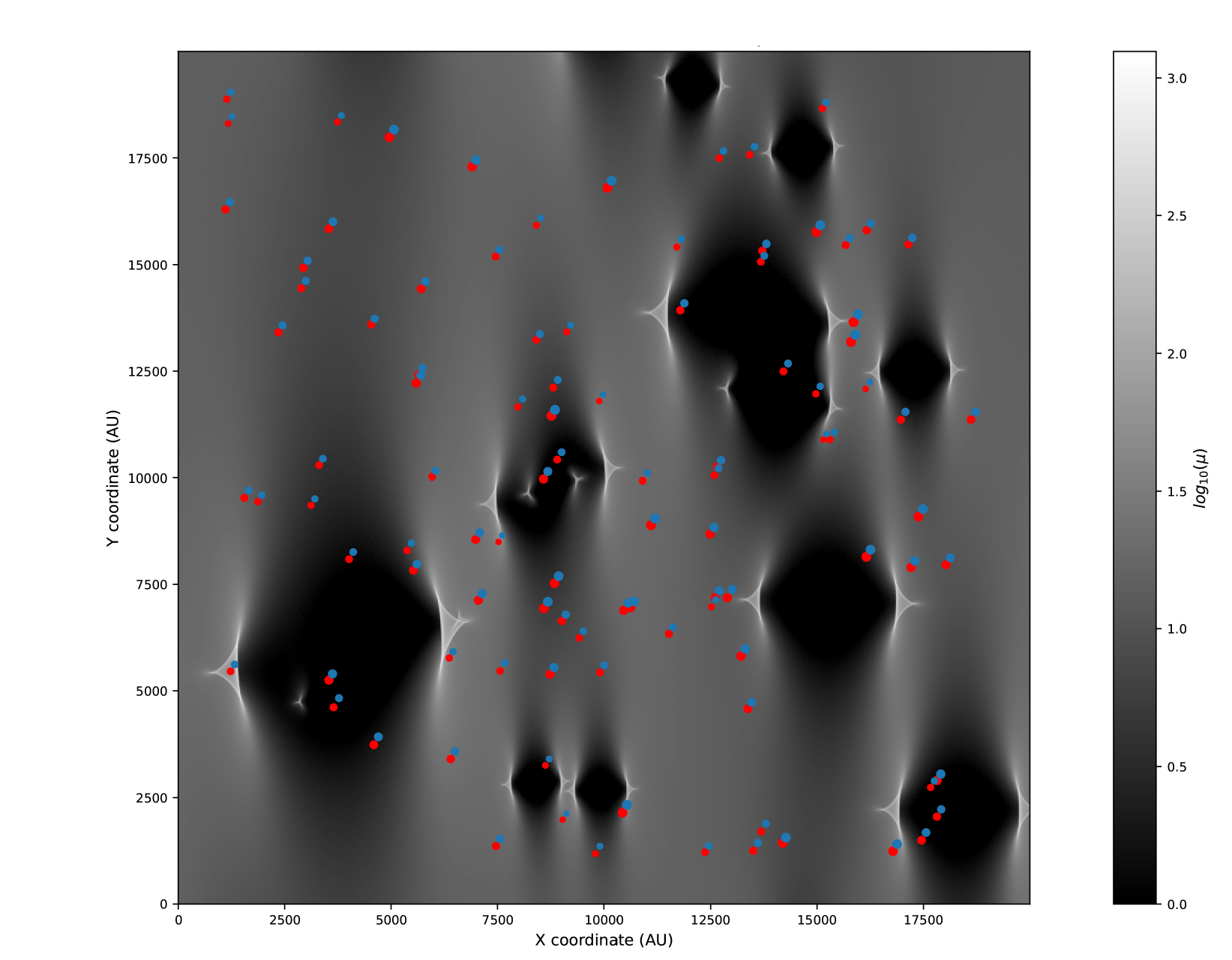}
    \caption{Illustration of stellar microlensing simulation described in Section~\ref{sec: simulation}, for which the background corresponds to lensing magnification (increasing from black to white) as indicated by the bar to the right.  Lensing is imposed by both A370 and intracluster stars, the latter generating microcaustics corresponding to regions of high magnifications (white regions).  Only a small fraction of the approximately $\sim$15 stellar microlenses in this field, corresponding to a surface mass density in intracluster stars of 50\,M$_{\odot}\,\textrm{pc}^{-2}$, is shown for the purpose of clarity.  The stellar microlenses are embedded in a macrolens (imposed by A370 over individual knots of the Dragon arc) having a macromagnification of $-13.5$.  Red dots represent individual stars belonging to a lensed star cluster in a background galaxy: these stars have been placed at random over the simulation field.  Blue dots represent the same stars seen 1\,yr later owing to a relative velocity of 1000\,km\,s$^{-1}$ at a position angle of $\sim$$30^{\circ}$ (measured clockwise from north) between A370 and the background lensed galaxy.  This relative motion changes (by either increasing or decreasing) the lensing magnification imposed on individual background stars, and may therefore change the overall brightness of their parent star cluster.
}
    \label{fig: microcaustics}
\end{figure*}

Figure~\ref{fig: microcaustics} shows a small region of our simulation area, over which the lensing magnification imposed on a star cluster is indicated in grayscale.  The lensing magnification over this area is composed of a macromagnification imposed by the galaxy cluster (the macrolens, as described by the lens model derived in Section\,\ref{sec: lens_model}), which imposes an essentially constant magnification over a background star cluster, as well as a web of micro-magnifications imposed by a collection of microlenses.  The latter can dramatically decrease or increase the lensing magnification imposed by the macrolens, thus creating localized regions of dark and white features in the magnification map that each corresponds to a microcaustic (associated with an individual microlens).  The typical scale of a microcaustic is a few thousand AU, and therefore orders of magnitude smaller than the typical size of a star cluster.

The color dots in Figure~\ref{fig: microcaustics} represent stars that have been randomly distributed over the simulation area.  The two different colors indicate their positions as observed at two different epochs owing to their motion relative to the microcaustics (in this case, from the blue to the red positions over time).  Their shifts in positions correspond to an assumed bulk velocity of $1000 \, \textrm{km\,s}^{-1}$ between the stars and the web of microcaustics (see details in Section~\ref{sec: velocity}).  At this velocity, a given star moves by 210\,AU relative to the microcaustics in one year (or 190\,AU between the two Flashlights observations separated by slightly less than a year).  As can be seen, although the macromagnification is constant over the simulation area, the motion of individual stars relative to the individual microcaustics can dramatically alter the brightness of individual stars,  and in this way potentially alter the collective brightness of the star cluster, thus seen as a lensed transient.

In summary, our simulation consists of four components: (i) the macrolens that contributes a constant magnification over the simulated region corresponding to an individual background star cluster; (ii) a collection of microlenses, which is characterized by their abundance or surface mass density; (iii) the mass (and therefore luminosity) function of the stellar population corresponding to the background star cluster; and (iv) a relative displacement between the stellar population and the web of microcaustics representing observations at two different epochs.  To understand how the individual components affect the result of our simulations, we explored a range of simulation parameters.  In the following subsections, the parameters of each component are described in more detail. 


\begin{deluxetable*}{c|c|c|c}
\label{tab: sim_param}
\centering
\tabletypesize{\scriptsize}
\tablewidth{0pt} 
\tablecaption{Tested Simulation Parameters}
\tablehead{Surface Mass Density, $\kappa_{\star}$ & Macromagnification, $\mu$ & Relative Displacement, $D$ (AU) & 30\,M$_{\odot}$ PBH fraction in DM, $f(M)$}
\startdata
0.005                 &   -7    & 20     &   0     \\
0.014*             &  -13.5*   & 100    &   1.5\% ($\kappa_{\star} = 0.012$)     \\
0.02                   &   -27    & 190*    &        \\
          &  & 400    &        \\
                 &       & 600    & \\
           &    & 1000   &    
\enddata
\tablecomments{Parameters denoted with * represent the physical condition of knot 2.6.2 in the two Flashlights observations}
\end{deluxetable*}

\subsubsection{Microcaustics}
\label{sec: muL_ICL}

In galaxy clusters, microlensing can be induced by intracluster stars or, in the situation considered here, stars at the outskirts of cluster members.  In either case, the stars involved are relatively old and therefore have relatively low masses.  We assume that these stars have a mass function described by a \citet{Kroupa} initial mass function (IMF) that is truncated at $M_{\odot} = 1.4$\,M$_{\odot}$, corresponding to the upper mass threshold for F-type stars.  Based on an analysis of the intracluster light performed by \citet{Morishita_2017}, we estimate the surface mass density of intracluster stars in the vicinity of the Dragon arc to be $\sim 10$\,M$_{\odot}\,\textrm{pc}^{-2}$ (equivalent to $\kappa_{\star} \approx 0.003$).  Knot 2.6.2 is located especially close to a cluster member (see Figure\,\ref{fig: RGB}), which makes a significant contribution to the stellar surface density at the position of this knot.  We compute a contribution from this cluster member of $\sim 40$\,M$_{\odot}\,\textrm{pc}^{-2}$ (equivalent to $\kappa_{\star} \approx 0.011$), and therefore a total stellar surface mass density at the position of knot 2.6.2 to be $\sim 50$\,M$_{\odot}\,\textrm{pc}^{-2}$ (equivalent to $\kappa_{\star} \approx 0.014$).  Details of this computation can be found in the Appendix.

On top of stellar microlenses, we also explore whether the addition of PBHs can better explain the observed transient rate among star clusters in the Dragon arc.  In our simulation, we consider PBHs having identical masses of 30\,M$_{\odot}$,  motivated by the LIGO detection of many merger events that indicate the possible existence of PBHs with such masses \citep{Jedamzik_2021}.  The result of our simulation, however, is not dependent on the exact mass of the PBHs within the range of 0.1--10\,M$_{\odot}$.

In each simulation run, we sample microlenses --- stars alone, or both stars and PBHs -- having a given surface mass density or combination of surface mass densities.  These microlenses generate microcritical curves on top of the macromagnification, $\mu$ (as given by our lens model for A370), in the image plane (at the redshift, $z = 0.375$, of the galaxy cluster).  When projected back to the source place (at the redshift, $z = 0.7251$, of the multiply-lensed galaxy that constitutes the Dragon arc) via the lens equation, these microlenses generate a web of microcaustics as shown in Figure\,\ref{fig: microcaustics}.  In our simulations, we use a pixel size of 1\,AU on a side in our computation box, driven by the need to well separate lensed stars belonging to star clusters in the Dragon arc, as well as to resolve changes in their positions between the two epochs of the Flashlight observations of A370 (owing to the motion of the lensed stars relative to the microcaustics).  This pixel size also ensures that each microcaustic (whether generated by stars or PBHs in the foreground cluster) is very well resolved, as can be visually appreciated from Figure\,\ref{fig: microcaustics}.  Our computation box has a size of $0.8\ \textrm{pc}\ \times\ 0.8\ \textrm{pc}$, in which we randomly distribute stellar microlenses having a number distribution with mass (i.e., mass function) drawn randomly from a Kroupa IMF truncated at 1.4\,M$_{\odot}$.  For a surface mass density of 50\, M$_{\odot}\,\textrm{pc}^{-2}$, our computation box would therefore contain $\sim 100$ stellar microlenses, ensuring a sampling in their mass function that is reflective of the adopted mass function given the relatively narrow mass range considered for these stars.

%
%

\subsubsection{Background Stellar Population}
\label{sec: IMF}

The lensed star clusters in the Dragon arc under consideration all have similar colors and SEDs, display emission lines, have rather similar macromagnifications according to our macrolens model, and have rather similar apparent brightnesses.  For simplicity, we therefore assume that all these star clusters have similar properties and, most importantly, the same stellar luminosity function.  Taking system 2.6 as being representative of these star clusters, we fitted a single stellar population to the SED of each of its lensed counterparts using {\tt Bagpipes} \citep{Carnall_2018}.  The best-fit model, for which we assume no dust extinction, is for a stellar population having an age of $\sim 4\,\textrm{Myr}$ and a total stellar mass $\sim10^{4.4}$\,M$_{\odot}$ (with a poorly constrained metallicity).  The estimated age is comparable to that of Westerlund\,1, and the estimated total stellar mass is nearly equal to the lower end of total stellar mass estimates for Westerlund\,1, the most massive star cluster known in our Galaxy, with an estimated mass spanning the range $\sim 10^{4.7}$--$10^{5.0}$\, M$_{\odot}$ \citep[see][and references therein]{Negueruela_2022}.  At such a young age, even relatively massive stars (late-0 and later spectral types) would still be on the main sequence, as required to produce an H\,II region detectable in emission lines.

As a starting point in our simulations, we assume a mass function for the lensed stars corresponding to a Kroupa IMF truncated at a mass of 100\, M$_\sun$, and covert their mass to luminosity using the prescription of \citet{Duric}.  As might be anticipated, and as we indeed find, any variations in the total brightness of a lensed star cluster between two epochs --- owing to changes in the positions of lensed stars relative to microcaustics --- are dominated by the most luminous stars (which contribute disproportionately to the luminosity of a star cluster).  In the Appendix, we truncate the mass function of the lensed stars at 10 \,M$_{\odot}$ (corresponding to mid-B stars) to investigate the consequences of a possible overrepresentation of luminous stars in our starting simulations.  Not surprisingly, the results differ for the two different upper limits in masses and therefore luminosities adopted for the lensed stars, but not by very much when considering only two observations separated by somewhat less than a year (as is the case for the Flashlights observations of A370).  When considering longer timescales, however, the results become increasingly divergent for the two mass cutoffs, as might be anticipated owing to a larger displacement between the lensed stars and microcaustics over a longer time interval.

As mentioned above, our computation box has a size of $0.8\ \textrm{pc}\ \times\ 0.8\ \textrm{pc}$, which is much smaller than the sizes of star-forming regions in our Galaxy having comparable total stellar masses \citep[e.g., Westerlund\,1, with an effective radius estimated at $\sim3.7\,\rm pc$;][]{Negueruela_2022}.  The size of our computation box is limited by the need to provide very high angular resolutions (pixel size of 1\,AU) so as to separate individual lensed stars as well as the displacement of these stars relative to the microcaustics owing to their relative motion separated by about a year, while maintaining a manageable computational time.  A single simulation of an entire star cluster would therefore require multiple computational boxes depending on the assumed sizes of the star clusters under consideration.  Recall from Section\,\ref{sec: muL_ICL} that the computational box contains $\sim 100$ stellar microlenses (depending on the surface mass density of intracluster stars under consideration) distributed at random over the computation box, and has a mass function drawn randomly from a Kroupa IMF truncated at $1.4\,\rm M_\sun$.  As the pattern of microcaustics in this computation box is representative of that in any other (indeed, this consideration is the main constraint on the minimal size of a computation box, as explored in more detail in the Appendix), rather than using multiple computations boxes (with a statistically similar pattern of microcaustics) for a single simulation of an entire star cluster, we simply place all the stars in the star cluster into one computation box. This approach is equivalent to combining (imagine superposing) multiple computation boxes, so long as the web of microcaustics in the combined computation box is representative of any single computation box; it also naturally lightens the computation load.

For a total stellar mass for the lensed star cluster of $10^{4.4}$\, M$_{\odot}$ and a mass function corresponding to a Kroupa IMF truncated at 100\,M$_\sun$, the total number of lensed stars in each computation box is therefore $\sim 54,000$.  In each simulation, we distribute these stars randomly over the computation box and impose a mass function drawn randomly from a Kroupa IMF truncated at 100\,M$_\sun$ (for which we draw a total of 5000 samplings in mass functions); to simulate their motion relative to the microcaustics as observed at two different epochs, all the stars are displaced by the same distance and at the same position angle (see Figure\,\ref{fig: microcaustics}). In this way, we performed 50,000 such simulations (i.e., using each of the 5000 samplings in mass functions ten times), with the lensed stars being displaced by the same distance but at a randomly selected position angle in each simulation.  To check whether we properly sample the most luminous stars, we find that every sampling in mass function (and hence every simulation) contains at least one star having a mass of $\ge 80$\, M$_\sun$, sometimes as many as twelve such stars, with a median number of three such stars among all these samples.

\subsubsection{Relative Displacement}
\label{sec: velocity}

The relative motion between the observer, microlenses (intracluster stars) contained within the macrolens (A370), and source (star clusters hosted by the multiply-lensed galaxy that forms the Dragon arc) can alter the total brightness of the source owing to changes in the microlensing magnification of its individual stars.  To assess the relative displacement between the microlenses and the source over a given time interval, we need to consider the relative transverse velocity involving the planes of the observer, lens, and source.  By taking into account factors including the peculiar velocity of  Earth and the host galaxy, the velocity dispersion of a merging galaxy cluster, and the motion of young star clusters with respect to their host galaxy, the relative transverse velocity is $
\sim 1000\,\textrm{km\,s}^{-1}$ \citep{Icarus}.  For a time interval of 330\,days between the two epochs of the Flashlights observations of A370, this relative transverse velocity translates to an observed transverse displacement of $\sim 190\,\textrm{AU}$ between the plane of the lens and the source.  Note that we have neglected the velocity dispersion of stars in the star cluster, for which the approximate velocity dispersion (given the inferred total stellar mass) is $\sim 2$ orders of magnitude smaller than the bulk velocity assumed above.

As mentioned in Section\,\ref{sec: IMF}, we displace the lensed stars along different (randomly selected) directions in different simulations, as we lack information on the angle of motion relative to the shear angle of the macrolens.  \citet{Dai_2021} has shown that varying the angle of motion relative to this shear angle does not significantly affect the brightness variation of a star cluster owing to microlensing.


\subsection{Simulation Results}
\label{sec: analytical}

\begin{figure}[ht]
    \centering
    \includegraphics[width =  \linewidth]{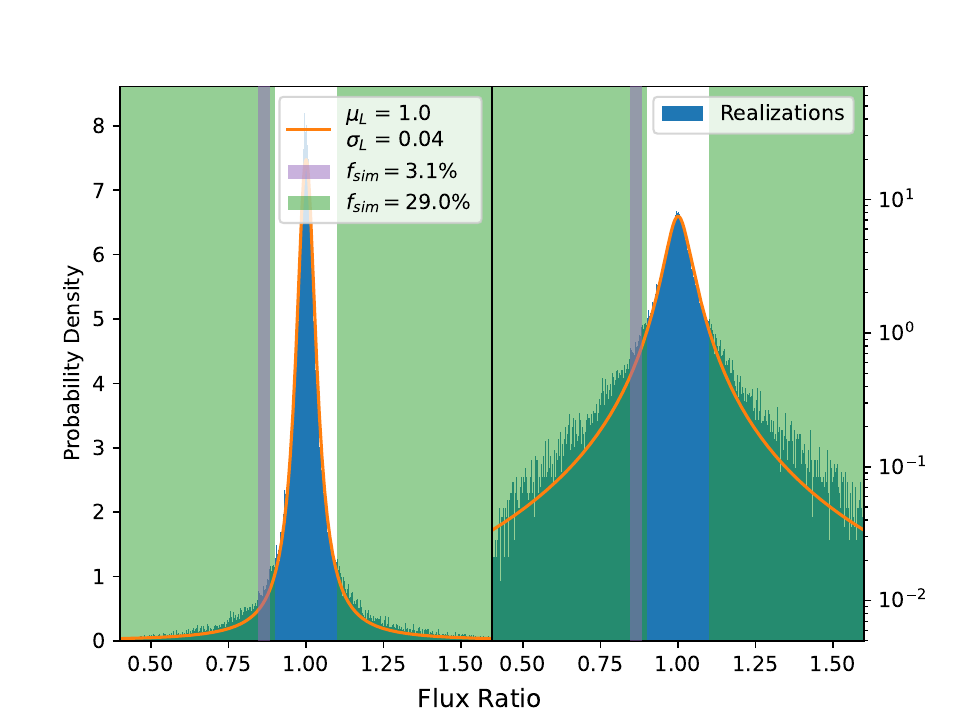}
    \caption{Probability of seeing, between two observations separated by 11\,months, an apparent brightness variation at a given level of a star cluster that is macrolensed by a galaxy cluster and microlensed by intracluster stars.  Probability distribution derived from a total of 50,000 different realisations of the simulation illustrated in Fig.\,\ref{fig: microcaustics}; in each realisation, stars belonging to the lensed star cluster were drawn at random from a Kroupa IMF truncated at $100 \, \rm M_{\odot}$, and placed at random over the simulation field.  All simulations employ a macrolens with a macromagnification of $\mu = -13.5$, a surface mass density in intracluster stars of $\Sigma_{\star} = 50$\,M$_{\odot}\,\textrm{pc}^{-2}$ ($\kappa_{\star} = 0.014$), and a relative displacement of $\sim$$190\, \textrm{AU}$ between the intracluster stars and lensed star cluster.  A Lorentzian function (orange curve) with a mean of $x_{0} = 1$ and a half width at half maximum of $\sigma_{L} = 0.04$ has been fitted to the probability distribution.  Left panel shows the probability distribution plotted on a linear scale, and right panel on a logarithmic scale to better visualise the tail of the Lorentzian distribution.  Green bands correspond to brightness variations $\ge 10\%$, for which the probability of seeing variations at such levels is 29.0\% (14,447 out of 50,000 realisations).  Purple band corresponds to the observed brightness variation of knot 2.6.2 given the $\pm 1\sigma$ measurement uncertainty in its brightness, for which the probability of seeing such a brightness variation is 3.1\% (1550 out of 50,000 realisations).}
    \label{fig: muL_hist}
\end{figure}

Figure\,\ref{fig: muL_hist} shows the probability distribution of brightness variations predicted by our simulations for knot 2.6.2, one of the lensed counterparts of a multiply-lensed star cluster in the Dragon arc, in observations separated by nearly 1\,yr (specifically, 330\,days).  As explained in Section\,\ref{sec: IMF}, this probability density is derived from 50,000 simulations in which, between two observations, changes in the positions of lensed background stars relative to a web of microcaustics --- generated by intracluster stars with a surface mass density commensurate with that estimated at the position of knot 2.6.2 (see Section\,\ref{sec: muL_ICL}) --- result in changes in the micromagnifications of the lensed stars.  These individual changes can add up to a collective change in their total brightness, especially when relatively scarce luminous stars experience relatively large changes in micromagnifications.

We find from Figure\,\ref{fig: muL_hist} that, most frequently, the cumulative effect of microlensing by intracluster stars results in no or imperceptible changes to the total brightness of a star cluster between one epoch and another separated by the time interval considered.  This result aligns with expectations that, most often, increases in micromagnifications for some stars are counterbalanced by decreases in micromagnifications for other stars, consistent with the findings by \citet{Dai_2021}.  Less frequently, this counteraction does not completely or nearly balance, resulting in larger changes --- either an increase or a decrease --- in the total brightness of a star cluster.  Much more rarely, one or more luminous stars experience an especially large change in micromagnification owing to changes in their positions relative to the microcaustics, therefore producing a correspondingly large change in the total brightness of a star cluster.

To obtain a useful semianalytical approximation to the results, we fit a Lorentzian function to the probability distribution shown in Figure\,\ref{fig: muL_hist} as given by 
\begin{equation}
    f(x; \, x_{0}, \sigma_{L}) = \frac{1}{\pi}(\frac{\sigma_{L}}{(x- x_{0})^{2} + \sigma_{L}^{2}})\, ,
\end{equation}
where $x$ is the total brightness of the star cluster, $x_0$ its average brightness (i.e., total brightness under macrolensing in the absence of microlensing), and $\sigma_{L}$ the scale parameter (i.e., half width at half-maximum intensity, HWHM) of the Lorentzian function.  In this way, the probability of seeing a change in total brightness of $x:x_0$ can be simply calculated provided knowledge of $\sigma_{L}$. The best-fit Lorentzian function to the probability density shown in Figure\,\ref{fig: muL_hist} is indicated by an orange curve.

\begin{figure*}[ht!]
    \centering
    \includegraphics[width =  \linewidth]{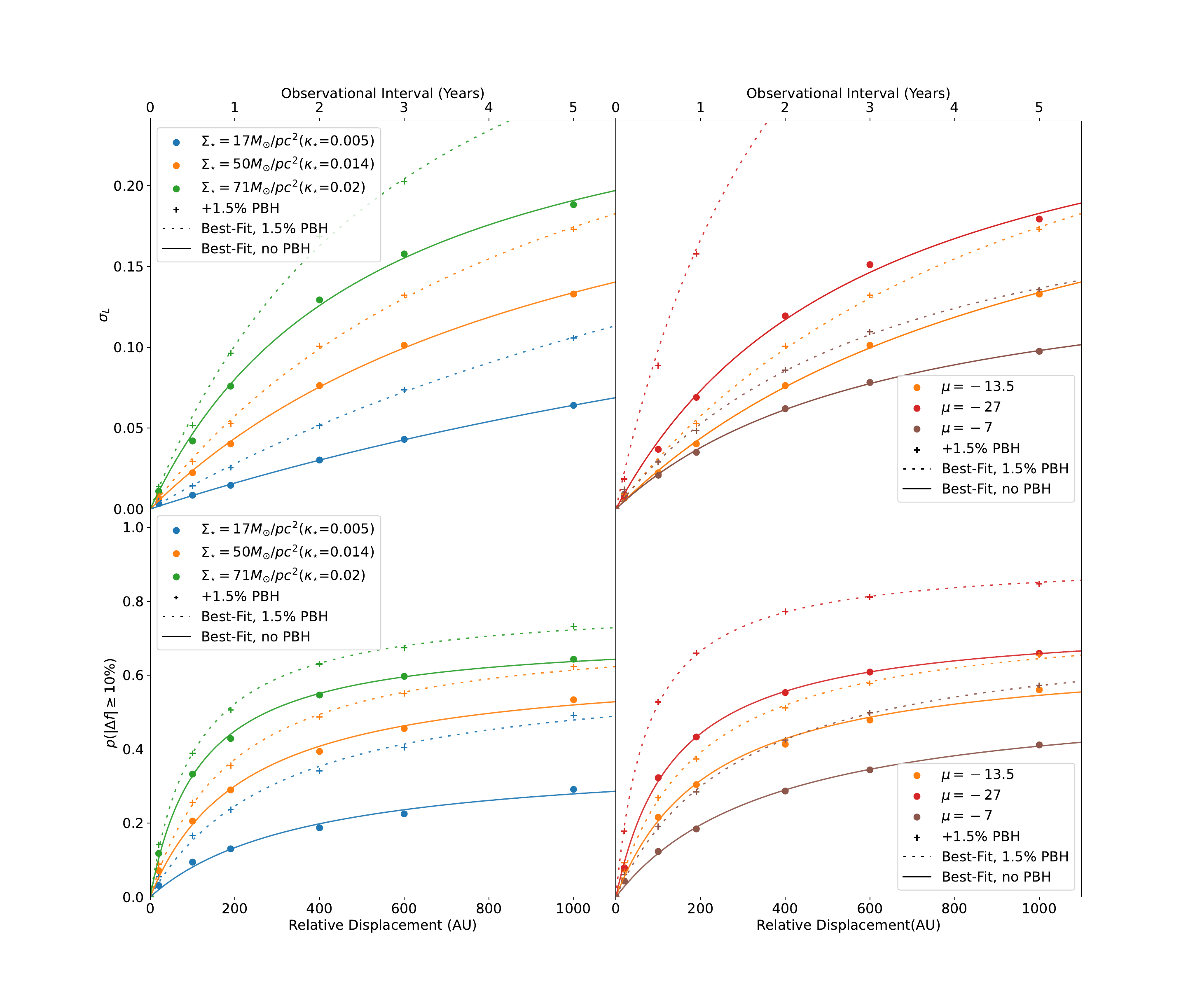}
    \caption{How probability distribution of Fig.\,\ref{fig: muL_hist} changes with surface mass density of microlenses, $\Sigma_{\star}$, at a given macromagnification, $\mu$ (left column), or with $\mu$ at a given $\Sigma_{\star}$ (right column), both as a function of different relative displacements (lower x-axis) between the microlenses and the lensed star cluster over different time intervals (upper x-axis) given a relative velocity of $1000 \rm \, km \, s^{-1}$.  Upper row shows half width at half maximum, $\sigma_{L}$, of Lorentzian functions fitted to probability distributions derived for different $\Sigma_{\star}$ at $\mu = -13.5$ (left panel), or different $\mu$ at $\Sigma_{\star} = 50 \,M _{\odot}\,\textrm{pc}^{-2}$ (equivalent to $\kappa_{\star} = 0.014$) (right panel), for the different values of $\mu$ and $\Sigma_{\star}$ specified in the legends.  Filled circles correspond to situations with no PBHs, for which solid curves represent best-fit Sigmoid functions to how $\sigma_{L}$ changes with either $\Sigma_{\star}$ or $\mu$.  Crosses correspond to situations where PBHs -- each having a mass of 30\,M$_{\odot}$ -- make up 1.5\% in total mass of the cluster-scale DM inferred for A370, for which dotted curves represent best-fit Sigmoid functions to how $\sigma_{L}$ changes with either $\Sigma_{\star}$ or $\mu$ in this situation.  Lower row shows corresponding probability of seeing a  brightness variation of at least 10\%, $p(|\Delta f| \ge 10\%)$, for which the curves now represent best-fit Sigmoid functions to how $p(|\Delta f| \ge 10\%)$ changes with either $\Sigma_{\star}$ or $u$ and when PBHs are added to the simulation.  
As is apparent, $p(|\Delta f| \ge 10\%)$ increases with increasing $\Sigma_{\star}$ at a fixed $\mu$, or increasing $\mu$ at a fixed $\Sigma_{\star}$, and when PBHs are added.  
}
    \label{fig: combined_motion}
\end{figure*}

In the upper row of Figure~\ref{fig: combined_motion}, we show how $\sigma_{L}$ changes with the (tangential) displacement, $D$, between stars and the microcaustics (lower $x$ axis) or, equivalently, the time interval between observations (upper $x$ axis).  The different color points show how our simulations predict $\sigma_{L}$ to change for different surface mass densities of intracluster stars, $\Sigma_{\star}$, at a given macromagnification (left panel), or for different macromagnifications, $\mu$, at a given $\Sigma_{\star}$ (right panel).  The color crosses show the corresponding predictions when we add PBH microlenses having identical individual masses of 30\,M$_{\sun}$ and totaling 1.5\% of the surface mass density of matter (dominated by DM), as predicted by our macrolens model, at the position of knot 2.6.2 (this surface mass density varying little over the region encompassing the Dragon arc).  As might be anticipated, irrespective of whether PBH microlenses are included, $\sigma_{L}$ becomes larger (the Lorentzian function broader) as $D$ increases, reflecting a larger possible change in the micromagnification of individual stars for a larger change in their positions relative to the microcaustics.  $\sigma_{L}$ also becomes larger as $\mu$ increases, as a given fractional change in micromagnification corresponds to a larger fractional change in stellar brightness for stars subject to larger macromagnification.  The dependence of $\sigma_{L}$ with either $\Sigma_{\star}$, including or excluding PBHs, or $\mu$ can be well approximated by Sigmoid functions, as indicated by the corresponding color curves, as expressed by
\begin{equation}
\label{eqn: power_law2}
\sigma_{L} \approx \frac{D}{K_{L} + a_{L}D} ,\quad D > 0 \, \, \, , 
\end{equation}
where $K_L$ and $a_L$ are the two constants of the Sigmoid functions.


In the lower row of Figure~\ref{fig: combined_motion}, we show how the probability of a total brightness change equal to or larger than 10\% of the mean, $p(|\Delta f| \ge 10\%)$, depends on $D$ for different $\Sigma_{\star}$ at a given $\mu$ (left panel), or for different $\mu$ at a given $\Sigma_{\star}$ (right panel).  We also show the corresponding dependencies when we add PBHs having a surface mass density as described above.  As might be anticipated, this probability becomes larger for a larger change in the positions of stars relative to the microcaustics, but up to a limit --- as the displacement approaches the typical separation between microcaustics, this probability approaches a constant value.  Just like for $\sigma_{L}$ as shown in the upper row of Figure~\ref{fig: combined_motion}, the dependence of $p(|\Delta f| \ge 10\%)$ with either $\Sigma_{\star}$, including or excluding PBHs, or $\mu$ can be well approximated by Sigmoid functions as expressed by
\begin{equation}    
\label{eqn: power_law}
p(|\Delta f| \ge 10\%) \approx \frac{D}{K_{p} + a_{p}D}\, ,\quad D > 0   
\end{equation}
where $K_p$ and $a_p$ are again the two constants of the individual Sigmoid functions.  The best-fit Sigmoid functions to our simulation results as indicated by the color points are indicated by the corresponding color curves.

To make Equations\,\ref{eqn: power_law2} and \ref{eqn: power_law} more generally applicable, first we normalize $D$ to a characteristic Einstein radius of $\theta_{\textrm{Ein}} = 210\, \textrm{AU}$, corresponding to the transverse displacement between the lensed stars and web of microcaustics over a year.  (We note that such a small Einstein radius is not physical for stars, as it would correspond to a microlens having $M \approx 0.0118$\,M$_{\odot}$ for a lens at $z = 0.375$ and a source at $z = 0.7251$, but chosen so as to provide a physically motivated normalization for $D$.)  This normalization permits us to express the transverse displacement between two observations as $\delta = D/\theta_{\textrm{Ein}}$.   
Second, as the best-fit Sigmoid functions are related to both the macromagnification ($\mu$) and the surface mass density of microlenses ($\Sigma_{\star}$ or $\Sigma_{\star} + \Sigma_{\rm PBH}$), with both combining to create the microcaustic network, we combine these two parameters by defining $\Sigma_{\textrm{eff}} = \mu \Sigma_{\star}$ or $\Sigma_{\textrm{eff}} = \mu (\Sigma_{\star} + \Sigma_{\rm PBH})$.  The latter is referred to as the effective surface mass density or the microlensing optical depth \citep{Diego_2018, Oguri_2018, Dai_2021, Palencia_23}.  Finally, to eliminate the effect of lensing geometry, we normalize $\Sigma_{\textrm{eff}}$ by the critical surface mass density,
\begin{equation}
    \Sigma_{\textrm{crit}} = \frac{c^{2}D_{s}}{4\pi G D_{ls}D_{l}}\, ,
\end{equation} 
\noindent where $c$ is the speed of light, $G$ the gravitational constant, and $D_{s}$, $D_{l}$, and $D_{ls}$ represent the distance to the source, the distance to the lens, and the distance between the source and the lens (respectively). Given the lens redshift of 0.375 and source redshift of 0.7251, the critical surface mass density is $\Sigma_{\textrm{crit}} = 3615$\, M$_{\odot}\,\textrm{pc}^{-2}$. This normalization converts the effective surface mass density into the form of convergence, $\kappa_{\textrm{eff}}$, thus allowing applications to situations with different critical surface mass densities.


\begin{figure}[ht!]
    \centering
    \includegraphics[width =  \linewidth]{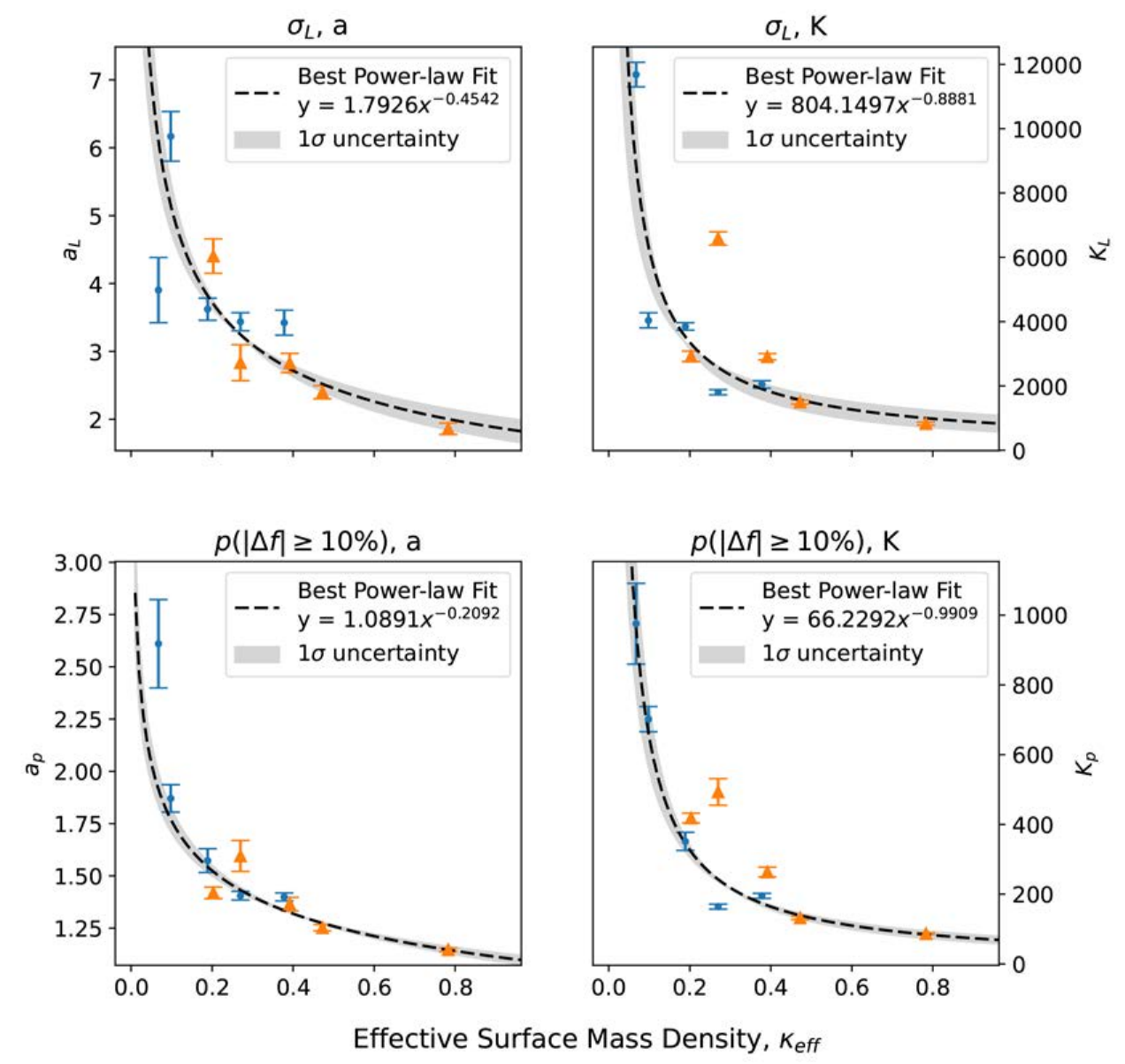}
    \caption{Upper row -- best-fit power laws to dependence of $a_L$ (left panel) and $K_L$ (right panel), the two constants of the sigmoid functions corresponding to the curves in the upper row of Fig.\,\ref{fig: combined_motion}, on $\kappa_{eff}$.   Lower row -- best-fit power laws to dependence of $a_p$ (left panel) and $K_p$ (right panel), the two constants of the sigmoid functions corresponding to the curves in the lower row of Fig.\,\ref{fig: combined_motion}, on $\kappa_{eff}$.  Best fit values are given in the legend for each panel.  Gray bands denote the $\pm 1\sigma$ uncertainty of the fit.  Blue circles are constants fitted to the curves in Fig.\,\ref{fig: combined_motion} that do not incorporate PBHs, and orange triangles are constants fitted to the curves in Fig.\,\ref{fig: combined_motion} incorporating PBHs that make up 1.5\% in total mass of the cluster-scale DM inferred for A370.  In this way, both $\sigma_L$ and $p(|\Delta f| \ge 10\%)$ can be expressed solely in terms of $\kappa_{eff}$ (see text).}
    \label{fig: power_law_fit_sigmoid}
\end{figure}

With these simplifications, the constants of the Sigmoid functions now depend solely on $\kappa_{\textrm{eff}}$ as shown in Figure~\ref{fig: power_law_fit_sigmoid}.  The blue data points correspond to the constants $a$ and $K$ of the different Sigmoid functions fitted to the simulation data points shown in Figure\,\ref{fig: combined_motion}.  We approximate the dependence of these two constants on $\kappa_{\textrm{eff}}$ by power-law functions, as indicated by the dashed curves in Figure~\ref{fig: power_law_fit_sigmoid} having the functional forms given in the legends, such that
\begin{equation}
\label{eqn: linear_fit}
K = S_{K} \times \kappa_{\textrm{eff}}^{\gamma_{K}}, \quad 0 \le \kappa_{\textrm{eff}} \le 1 \, ,
\end{equation}
\begin{equation}
\label{eqn: linear_fit2}
a = S_{a} \times \kappa_{\textrm{eff}}^{\gamma_{a}}, \quad 0 \le \kappa_{\textrm{eff}} \le 1 \, \, \, ,
\end{equation}
\noindent where $S$ and $\gamma$ are the coefficients and the exponents (respectively) of the power-law fits, and have subscripts denoting whether they correspond to the constants $K$ or $a$ of the Sigmoid function.

\begin{figure*}[ht!]
    \centering
    \includegraphics[width = 1 \linewidth]{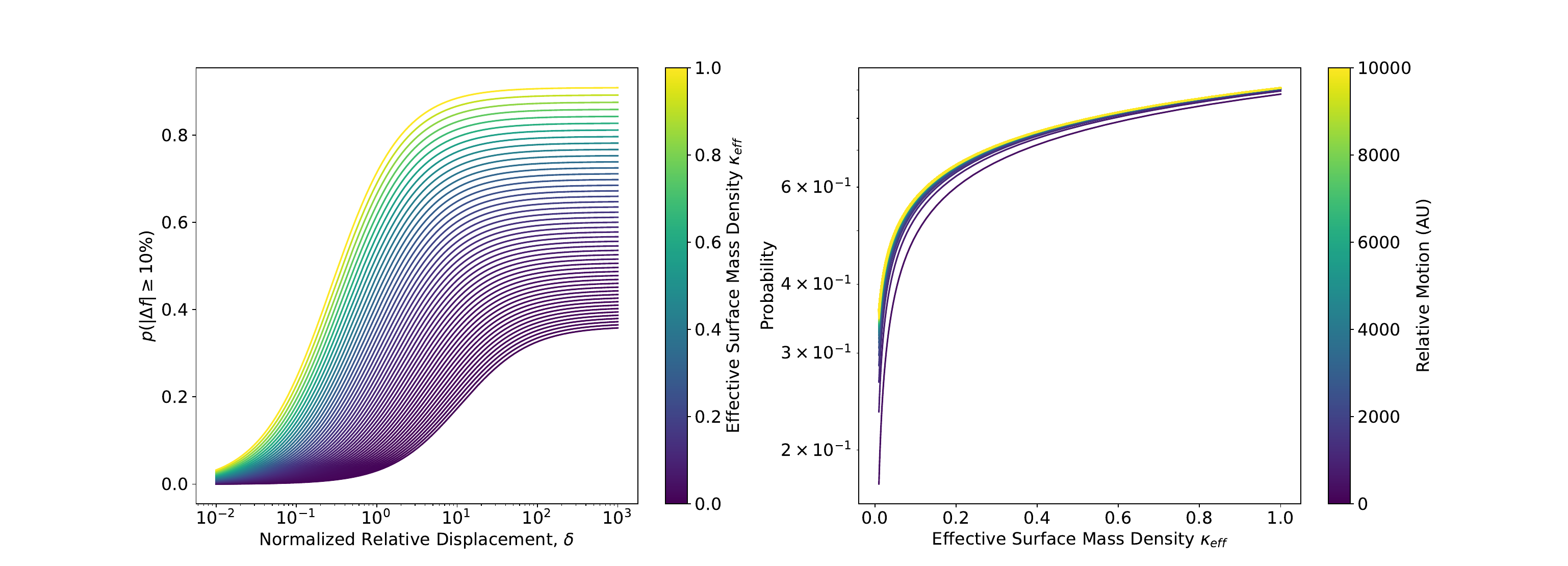}
    \caption{Left panel showing dependence of $p(|\Delta f| \ge 10\%)$ on $\delta$ (related to the time interval between observations) at different $\kappa_{\textrm{eff}}$ (related to the surface mass density of microlenses for a given macrolens).  Right panel showing dependence of $p(|\Delta f| \ge 10\%)$ on $\kappa_{\textrm{eff}}$ at different $\delta$.  A detailed discussion of the different dependencies is provided in the Appendix.}
    \label{fig: pvD}
\end{figure*}

By substituting the best-fit coefficients back to the Sigmoid functions and rearranging the terms, the dependence of $\sigma_{L}$ and $p(|\Delta f| \ge 10\%)$ on both $\delta = D/\theta_{\textrm{Ein}}$ and $\kappa_{eff} = \Sigma_{\textrm{eff}}/\Sigma_{\textrm{crit}}$ can simply be expressed as

\begin{equation}
\label{eqn: power_law_fitted2}
\sigma_{L} \approx \frac{\kappa_{\textrm{eff}}^{0.9}}{3.8\delta + 1.8\kappa_{\textrm{eff}}^{0.45}}\, ,
\end{equation}

\begin{equation}
\label{eqn: power_law_fitted1}
p(|\Delta f| \ge 10\%) \approx \frac{\kappa_{\textrm{eff}}}{0.3\delta^{-1}+1.1\kappa_{\textrm{eff}}^{0.8}} \, ,
\end{equation} 
\noindent for which both parameters scale with $\kappa_{\textrm{eff}}$ to first order.  These equations can be used to estimate both $\sigma_{L}$ and $p(|\Delta f| \ge 10\%)$ by simply substituting different values of $\delta$ and $\kappa_{\textrm{eff}}$ under the boundary condition $\delta > 0$ and $0 \le \kappa_{\textrm{eff}} \le 1$.  For example, the dependence of $p(|\Delta f| \ge 10\%)$ with $\delta$ (related to the time interval between observations) at different $\kappa_{\textrm{eff}}$ (related to the surface mass density of microlenses for a given macrolens) is shown in Figure~\ref{fig: pvD}.  For reasons already mentioned above, the probability of a brightness variation of this magnitude increases as the displacement between the lensed stars and microlenses (and hence the time interval between observations) increases, and as the surface mass density of microlenses increases at a given macromagnification.  Other trends apparent in this figure are explored and explained in the Appendix.


\subsection{Comparison with Observations}
\label{sec: muL_explain}

We compare the predictions of our simulations with the Flashlights observations of the Dragon arc.  Specifically, could all the (relevant) transients in the Dragon arc be induced by microlensing of star clusters such as to alter their individual collective brightness between two observations?  Are PBHs required in addition to intracluster stars, along with stars associated with clusters members where relevant, to explain the observed transient detection rate?  We begin our assessment by considering just system 2.6 as described in Section\,\ref{subsect_6p2}.  We then extend our assessment to all the young star clusters in the Dragon arc --- regardless of whether they display brightness variations --- that ought to be subjected to stellar microlensing under the conditions considered in our simulations, as described and explained in more detail in Section\,\ref{sec: detection_rate}.

\subsubsection{System 2.6}\label{subsect_6p2}

Among the three lensed counterparts in System 2.6, knot 2.6.2 is subjected to the highest surface mass density of stellar microlenses ($\kappa_{\star}$) owing to its proximity to a neighboring cluster member.  Furthermore, according to our macrolens model for A370, knot 2.6.2 is subjected to the highest macromagnification ($\mu$).  Having therefore the largest $\kappa_{\textrm{eff}} = \mu \times \kappa_{\star} \approx 13.5 \times 0.014 = 0.189$, knot 2.6.2 is predicted to have the highest probability of displaying --- owing to stellar microlensing --- the largest brightness variations among the three lensed counterparts of system 2.6.  Indeed, among these counterparts, knot 2.6.2 displays the largest brightness variation between the two epochs of the Flashlight observations, in qualitative agreement with the predictions of our simulations.


Although we detected brightness variations for both knots 2.6.2 and 2.6.3 between the two epochs of the Flashlights observations, only knot 2.6.2 displayed a brightness variation greater than 10\%.  We now calculate the probability of a brightness variation greater than 10\%, $p(|\Delta f| \ge 10\%)$, among the three lensed counterparts of system 2.6 for a lateral displacement between the stellar microlenses and the background star cluster of $\delta = 0.9$, corresponding to the time interval between the two Flashlights observations.  Both knots 2.6.1 and 2.6.3 are subjected to a similar surface mass density of stellar microlenses and a similar macromagnification, such that $\kappa_{\textrm{eff}} \approx 10 \times 0.003 = 0.03$ for each knot.  Our simulations therefore predict $p(|\Delta f| \ge 10\%) = 7.2\%$ for knot 2.6.1 and knot 2.6.3 individually.  By comparison, for knot 2.6.2, $p(|\Delta f| \ge 10\%) = 29.6\%$.   The joint probability that only knot 2.6.2 displays a brightness variation greater than 10\% is therefore $29.6\% \times (1-7.2\%)^2 = 25.5\%$: thus, our simulation predicts a one in four chance that knot 2.6.2 displays a brightness variation of the magnitude observed between the two epochs of the Flashlights observations.  That we observe such a brightness variation for knot 2.6.2 but not for the other lensed counterparts is therefore not surprising, although more observations are of course needed for a meaningful statistical test.

For completeness, we can also compute the probability of a brightness variation of the magnitude observed for knot 2.6.2 over a time interval corresponding to that between the two Flashlights observations.  From the simulations specifically tailored to this knot ($\kappa_{\star} = 0.014$, $\mu = -13.5$, and $\delta = 0.9$), we find that 1550 of the 50,000 realizations fall within the observed range in brightness change for knot 2.6.2 of $0.86 \pm 0.02$ ---  a 3.1\% probability of a brightness variation at the magnitude observed for knot 2.6.2.  Thus, a brightness variation of the magnitude detected is unlikely to be repeated in future observations of this knot.


\subsubsection{Transient Detection Rate}
\label{sec: detection_rate}

The many transients detected in the Dragon arc over the two epochs of the Flashlights observations can provide a more meaningful statistical test of microlensing.  To select systems relevant to our simulations, we note that transients located at or near the critical curve of the macrolens are subjected to such high macromagnifications that their brightness variations can potentially be induced by changes in the microlensing of just a single background lensed star, rather than the collective effects of microlensing on a background stellar population as considered in our simulations.

To omit transients that potentially correspond to microlensing of just a single highly-magnified background star, we impose the following conditions when selecting star clusters in the Dragon arc to compare with our simulations.
\begin{enumerate}
    \item Located well away from the critical curve of the macrolens, thus selecting only knots having $\mu_{\textrm{macro}} \le 50$ according to our macrolens model for A370.
    \item Detectable in both Flashlights epochs, to mitigate against transients corresponding to changes in microlensing of a single background lensed star. 
    \item $\textrm{FWHM} \le 0.08\arcsec$ such that they are unresolved, to avoid difficulty in the photometry of extended knots (especially those at the eastern extreme, corresponding to the ``head" of the Dragon arc).
\end{enumerate}
In this way, we select a total of 55 knots.  All these knots have relatively blue colors (and therefore likely correspond to young star clusters), and are associated with line-emitting regions at the angular resolution of the MUSE observations --- although a single spatial element of the MUSE maps can contain more than one knot, so that we cannot be certain that every single knot selected is still sufficiently young to be associated with H\,II regions.  Among these 55 knots, we detect brightness variations greater than 10\% among 8 knots, yielding an observed transient detection rate of $14.5 \pm 5.1\%$ (the uncertainty reflecting Poisson statistics).  

For simplicity in estimating the predicted transient detection rate, we assume that all the knots have similar stellar luminosity functions and total stellar masses, corresponding to that adopted for knot 2.6.2.  According to our macrolens model, the typical macromagnification of the selected knots (i.e., averaged over all these knots) is $\mu \approx 20$.  None except knot 2.6.2 is near a cluster member, and so we adopt a surface mass density for stellar microlenses of $\kappa_* \approx 0.003$ (see Section~\ref{sec: muL_ICL}), corresponding to that estimated for intracluster stars alone at the region of the Dragon arc.  Substituting $\kappa_{\textrm{eff}} = 20 \times 0.003 = 0.06$ and $\delta = 0.9$ into Equation~\ref{eqn: power_law_fitted1}, our simulations predict $p(|\Delta f| \ge 10\%) = 12.9\%$ among these knots.  The predicted rate is in excellent agreement with the observed transient detection rate of $p(|\Delta f| \ge 10\%) = 14.5 \pm 5.1\%$ as mentioned above.  This agreement suggests that stellar microlensing alone can explain the transients observed among the star clusters in the Dragon arc located well away from the critical curve of the macrolens: there is no need to invoke additional microlensing from a population of PBHs having masses of 0.1--10\,M$_\sun$, or of intrinsic variability associated with especially luminous stars.  The upper limit placed on the mass fraction of DM in A370 that could conceivably be composed of such PBHs will be deferred to a future paper.

To address the likelihood that any single of the aforementioned transients is induced by stellar microlensing, we calculate the probability of such an event using Bayesian statistics,
\begin{equation}
P(\mu L| Tr) = \frac{P(Tr|\mu L)P(\mu L)}{P(Tr)}\, ,
\label{eqn: Bayesian}
\end{equation}
where $\mu L$ denotes the predicted microlensing rate for a given change in brightness, and $Tr$ the observed transient detection rate for the same change in brightness. As described above, the predicted microlensing detection rate having brightness variations greater than 10\% is $P(Tr|\mu L) = 12.9\%$, compared with the observed transient detection rate at the same level of brightness variation of $P(Tr) = 14.5\% \pm 5.1\%$.  As microlensing always acts on lensed knots, even when no brightness variations are detectable between two observations, $P(\mu L) = 100\%$.  Combining all these rates together yields a probability of $P(\mu L|Tr) = 89^{+11}_{-31}\%$ for any single of the aforementioned transients to be induced by stellar microlensing.  Hence, our simulations predict that all these transients are almost certainly induced by stellar microlensing.

Note that we only consider the effects of microlensing on lensed images with negative parities, where $\mu_{t} < 0$ (i.e., inwards of the critical curve associated with the macrolens).   The effect of microlensing -- and, in general, small-scale structures -- is more pronounced on images with negative than positive parities \citep[e.g., ][]{Oguri_2018, williams2023flashlights, Palencia_23}, whereby $\mu_{t} > 0$ (i.e., outwards of the critical curve associated with the macrolens).  Forty of the fifty-five young star clusters we identify in the Dragon arc are positive-parity images, such that half of these positive-parity images are located at the Dragon's ``head'' and most are not found to be multiply lensed.  We find five transients among the 40 positive-parity images (detection rate of $12.5 \pm 5.6\%$), compared with three transients among the 15 negative-parity images (detection rate of $20 \pm 11.5\%$).  The roughly comparable detection rates motivate future work to consider the effects of microlensing on lensed images with positive parities, and may hint at other contributors to lensed transients.

\section{Discussion}  \label{sec: Discussion}

Throughout our description thus far, we have ignored additional lensing contributions by another type of intracluster object: globular clusters (GCs).  In the case of knot 2.6.2, located close to a cluster member, a GC bound to that cluster member could elevate the lensing magnification at knot 2.6.2 if this GC happens to lie along the sightline towards this knot.  Here, we briefly consider how GCs, whether they be bound to cluster members or tidally stripped from cluster members to freely roam the ICM, may alter the results obtained in Section\,\ref{sec: Extrinsic}.

\begin{figure}[ht!]
    \centering
    \includegraphics[width = 0.7 \linewidth]{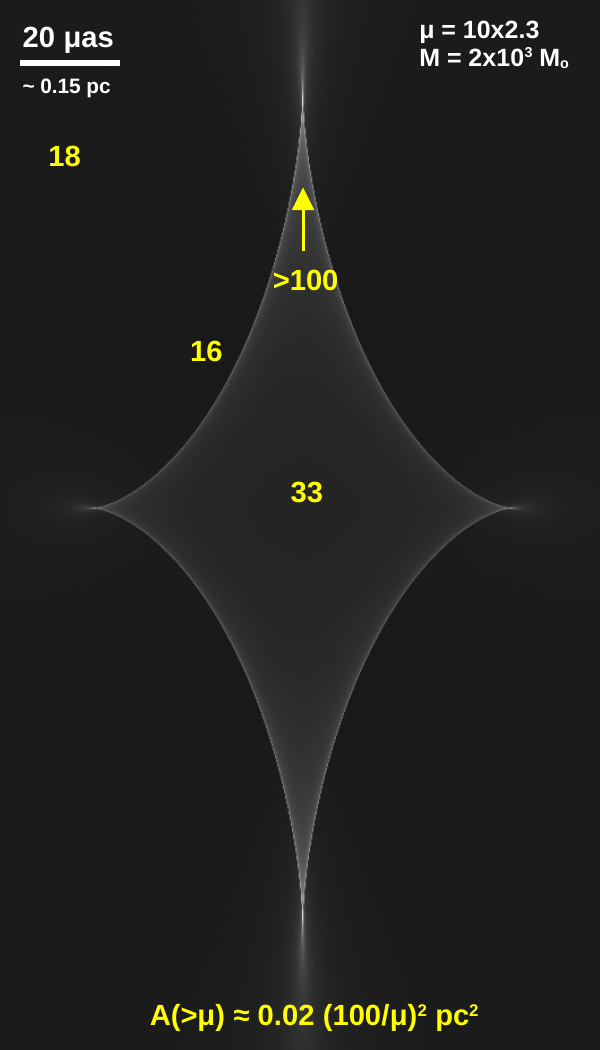}
    \caption{Millicaustic around a millilens with a mass of $M=2\times10^3\, {\rm M}_{\odot}$ (i.e., comparable to a low-mass globular cluster) in a region of the lens plane where the macromagnification of 23.  Numbers in yellow indicate the typical magnifications at the respective locations.  The area with magnification greater than 100 is very small ($\sim 0.02$\, pc$^2$) and concentrated around the thin caustic lines and cusps.  For a macromagnification of $13$, the area would be a factor of $(23/13)^2$  smaller.}
    \label{fig: Millilens}
\end{figure}

Figure~\ref{fig: Millilens} shows the caustic, at the redshift of the multiply-lensed galaxy corresponding to the Dragon arc, associated with a GC of mass $M = 2 \times10^3 \,{\rm M}_{\odot}$ if located in a region of A370 where the macrolensing magnification is 23.  This situation quite closely mimics the effect of a GC that happens to lie along the sightline to knot 2.6.2.  For the adopted GC mass and macrolensing magnification, the lateral scale of the caustic is $\sim$1\,pc.  Larger GC masses or larger macrolensing magnifications would result in even larger physical scales for the caustic; nonetheless, much of the region enclosed by or immediately beyond the caustic is only moderately elevated in magnification (to within a factor 2 of the macrolensing magnification).  Only regions very close to the thin caustic lines or the somewhat larger regions around the cusps of the caustic have highly elevated magnifications.  In these regions, however, stars associated with the lensed background star cluster can be highly magnified, thus further boosting the effect of microlensing by intracluster stars or PBHs.  

For instance, at a combined lensing magnification for the GC and macrolens of $\mu>100$, $\kappa_{\textrm{eff}}$ is sufficiently large that the probability of microlensing inducing a change in the total brightness of a lensed background star cluster at a given level between two observations becomes highly elevated.  We now consider the area of the background star cluster over which the combined lensing magnification imposed by the GC and macrolens is so high that stars in this area have an outsized effect on the total brightness of the star cluster owing to microlensing.  For the GC under consideration, the area over which $\mu > 50$ scales as $A(>\mu) \approx 0.02 (100/\mu)^2$\, pc$^2$.  Assuming that the lensed background star cluster spans tens of parsecs, only a small area in total of the star cluster is so highly elevated in magnification that microlensing plays an outsized role in altering its total brightness; i.e., $A(\mu \ge 100) \approx 0.02$\, pc$^2$.  
For GCs having larger masses, however, the effect can be much greater, as the area over which the lensing magnification of the GC and macrolens combined exceeds $\mu \approx 50$ scales linearly as the GC mass \citep{Diego_2018,Palencia_23,Diego2024_3M}.  For GCs having $M \approx 10^5 \,{\rm M}_{\odot}$, at the peak of the GC mass function for galaxies in the local universe, $A(\mu \ge 100) \approx 2$\, pc$^2$, further elevating the number of stars in a star cluster that can have an outsized effect on the total brightness of a star cluster owing to microlensing.  As the statistics for lensed transients associated with star clusters improve, future work will need to fold in (intracluster) GCs to study their effects on the detection rate of such transients.

\section{Summary and Conclusions}
\label{sec: conclusions}
As part of the Flashlights program, we acquired deep images of Abell 370 (A370) with the {\it HST} through the UVIS F200LP filter, which spans the broad wavelength range 1990--10550\,\AA.  These images were taken at two epochs separated by nearly 1\,yr (330\,days), and at the same telescope roll angle so that the imprints left by the telescope point-spread-function remain the same in both images.  Differencing these two images, we discovered numerous spatially-unresolved brightness variations in the Dragon arc (see Figure\,\ref{fig: transient_detection}), comprising a late-type galaxy at $z = 0.7251$ that is multiply lensed by the galaxy cluster A370 at $z = 0.375$.  Some of these lensed transients are located near critical curves, and not associated with any apparent persistent features.  Such transients have previously been attributed to changes in lensing magnification imposed on an individual background luminous star as it moves across the sky relative to intracluster stars in the foreground galaxy cluster: as the background star moves across a caustic associated with the foreground stellar microlens and the cluster macrolens combined, it brightens dramatically before fading to obscurity.

In this paper, our focus is on a different type of lensed transient in the Dragon arc: those associated with persistent knots.  We identify these knots as young star clusters based on their colors and spectra, all of which display emission lines characteristic of H\,II regions.  Their brightness variations may constitute either dimming or brightening from their commonly observed brightness levels based on other images of A370 taken in the Hubble Frontiers Fields (HFF) and Beyond Ultra-deep Frontier Fields And Legacy Observations (BUFFALO) programs.   From just the two Flashlights images alone, we discovered 10 such transients in the Dragon arc (enclosed by circles in Figure\,\ref{fig: transient_detection}) having brightness variations $\ge 3\sigma$, where $\sigma$ is the measurement uncertainty of their individual brightnesses.  

What generates these newly-discovered types of lensed transients?  We considered two possibilities: (i) intrinsic variability associated with an energetic outburst on a single luminous star in the star cluster; and (ii) imposed extrinsically on all the stars in the background star cluster through collective microlensing by a population of intracluster stars acting alone or in concert with primordial black holes (PBHs).  

Two of the transients of interest, displaying comparable variations (dimming) in brightnesses between the two Flashlights epochs, are associated with two of the lensed counterparts of a triply-lensed star cluster (knots 2.6.2 and 2.6.3, which along with knot 2.6.1 are indicated by arrows in Fig.\,\ref{fig: transient_detection}).  This situation permits a test of intrinsic variability, in which case a brightness variation in one lensed counterpart ought to be repeated in another lensed counterpart albeit with a time difference owing to their different light arrival times at the observer.  For these particular transients, we found the following:

\begin{enumerate}
        
    \item  If associated with an outburst from a Luminous Blue Variable (LBV), then the lensing magnification imposed by the cluster macrolens is sufficient to produce the level of brightness variation observed for a given lensed counterpart.   
    
    \item  Our lens model for A370 predicts significantly different intrinsic brightnesses for lensed counterparts of the same star cluster in the Dragon arc, even those having lensed counterparts that show no detectable brightness variations between the two Flashlights epochs (see Figure\,\ref{fig: LC_all}).  These small differences reflect the limited accuracy at which our lens model is able to predict lensing magnifications over the Dragon arc.  As a consequence of this limited accuracy, we are not able to make a conclusive test of intrinsic variability based only on images taken at two epochs -- as the light from LBV outbursts may display more complex time variability than just a simple monotonic dimming.  Nonetheless, this test remains feasible once images taken over more epochs become available, allowing us to test whether the pattern of light variability in one lensed counterpart is repeated in the others at different observed times.
            
\end{enumerate}

To test for extrinsic variability, we conducted simulations in which we superposed a population of intracluster stars, alone or in concert with PBHs, on a population of background stars associated with a star cluster (see Fig.\,\ref{fig: microcaustics}).  The purpose of these simulations is to test how much the total brightness of a star cluster can change between two epochs owing to the motion of the background star cluster relative to the foreground microlenses.  Specifically, we computed the probability of a given fractional change in brightness -- either increase or decrease -- as a function of the time interval between two observations (see Fig.\,\ref{fig: pvD}).  In the simulations, we assumed a mass function for the intracluster stars corresponding to the Kroupa initial mass function (IMF) truncated at $1.4 \, M_\sun$.  We adopted a range of surface mass densities for these stars around the value inferred at the vicinity of the Dragon arc.   For the stellar population in the lensed background star cluster, recalling that the ones selected for study are relatively blue and sufficiently young to still exhibit H\,II regions, we assumed a mass function corresponding to the Kroupa initial mass function (IMF) truncated at $100 \, M_\sun$.  We also explored how truncating the mass function of the star cluster at $20 \, M_\sun$ in the Appendix, thereby changing the maximal luminosity of their constituent stars, affects the results.  From these simulations, we found the following:

\begin{enumerate}
    \item  Over two epochs separated by somewhat less than 1\,yr, the probability of a given brightness variation only changes slightly (by $\sim$10\%) for the two different cutoffs in maximal stellar luminosities in the lensed background star cluster.  Over epochs separated by longer time intervals, however, the results increasingly diverge as the background stars experience a greater lateral displacement relative to the web of microcaustics associated with the microlenses and the cluster macrolens combined. 
        
    \item  Selecting only transients that show a brightness variation of $\ge 10\%$ among all the young star clusters identified in the Dragon arc, irrespective of whether these star clusters exhibit brightness variation among any of their lensed counterparts, the observed transient rate is compatible with those predicted by our simulations employing only intracluster stars acting alone as the foreground microlenses.  The upper limits placed on the fractional mass in Dark Matter (DM) that might possibly constitute PBHs is deferred to a companion paper.
    
\end{enumerate}

Finally, we noted that globular clusters (GCs), whether bound to cluster members or tidally stripped from cluster members to freely roam the ICM, can have an appreciable effect on our results if located along the sightline to a lensed background star cluster.  Although highly elevating the lensing magnifications over only a relatively small region of the lensed background star cluster, stars in this region can be so highly magnified as to elevate the probability of microlensing inducing a change in the total brightness of their parent star cluster at a given level between two observations.  As the statistics for lensed transients associated with star clusters improve, future work will need to fold in (intracluster) GCs to study their effects on the detection rate of such transients.  

Further improvements in observations of lensed transients can be provided by deep and multi-band observations of galaxy clusters at closely-spaced epochs so as to examine any changes in colors.  Such color changes may reflect intrinsic variability associated with a single highly-magnified star such as a luminous blue variable (LBV), or associated with the subset of stars that contribute most to changes in brightness of a star cluster between two epochs owing to the collective effects of microlensing.  Such images may soon become available from observations with the {\it JWST}.

\section*{Acknowledgement}

We thank Juno Li, Arsen Levitskiy, and Josh Zhang for the many useful discussions for this work. 

S.K.L., J.L., T.B., and A.C. acknowledge the Collaborative Research Fund under grant C6017-20G, which is issued by the Research Grants Council of Hong Kong SAR.
J.M.D. acknowledges the support of project PID2022-138896NB-C51 (MCIU/AEI/MINECO/FEDER, UE) Ministerio de Ciencia, Investigaci\'on y Universidades. 
P.L.K acknowledges NSF grant AAG 2308051.
J. L. acknowledges support from the Research Grants Council (RGC) of Hong Kong through the General Research Fund (GRF) 17312122.
A.A. acknowledge the support of a General Research Fund (GRF) under grant 17312122 administered by the RGC, as well as a Dissertation Year Fellowship (DYF) from the University of Hong Kong.
A.K.M. and A.Z. are grateful for grant 2020750 from the United States--Israel Binational Science Foundation (BSF) and grant  2109066 from the U.S. National Science Foundation (NSF); by the Ministry of Science \& Technology, Israel; and by the Israel Science Foundation grant 864/23.
A.V.F. is supported by the Christopher R. Redlich Fund and many individual donors. 

This research is based on observations made with the NASA/ESA {\it Hubble Space Telescope} obtained from the Space Telescope Science Institute, which is operated by the Association of Universities for Research in Astronomy, Inc., under NASA contract NAS 5–26555. These observations are associated with programs GO-14038, 15117, 15936, and 16278. We acknowledge support from NASA programs GO-15936, GO-16278, GO-16729, and GO-17504.

We acknowledge the usage of the following programs: Astropy \citep{astropy:2013, astropy:2018}, Numpy \citep{numpy}, Matplotlib \citep{matplotlib} and Scipy \citep{scipy}.

\bibliography{sample631}{}
\bibliographystyle{aasjournal}

\appendix

\section{Multiply-Lensed Image Systems}
\label{sec: system_list}

The sky positions and redshifts of the multiply-lensed systems used to constrain our lens model for the Dragon arc, referred to in Section~\ref{sec: lens_model}, are given in Table~\ref{tab: System_list}.  All redshifts are determined spectroscopically except for those denoted by asterisks, for which their redshifts are determined either by fitting model spectral templates to SEDs or through lensing geometry.

\startlongtable
\begin{deluxetable}{llll}
\label{tab: system_list}
\centering
\tabletypesize{\scriptsize}
\tablewidth{0pt} 
\tablecaption{List of Multiply-Lensed Images in A370}
\tablehead{ID   & R.A.        & Dec        & Redshift }
\startdata
\hline
1.1  & 39.976273 & -1.5760558 & 0.8041   \\
1.2  & 39.968691 & -1.5766113 & 0.8041   \\
1.3  & 39.967047 & -1.5769172 & 0.8041   \\
\hline
2.1.1  & 39.973825 & -1.584229  & 0.7251   \\
2.1.2  & 39.971003 & -1.5850422 & 0.7251   \\
2.1.3  & 39.96963  & -1.5848508 & 0.7251   \\
2.1.4  & 39.969394 & -1.5847328 & 0.7251   \\
2.1.5  & 39.968722 & -1.5845058 & 0.7251   \\
2.2 - 2.10 (1)& -- & -- & 0.7251\\
\hline
3.1  & 39.978925 & -1.5674624  & 1.9553   \\
3.2  & 39.968526 & -1.5657906 & 1.9553   \\
3.3  & 39.965658 & -1.566856 & 1.9553   \\
\hline
4.1  & 39.979704 & -1.5764364 & 1.2728   \\
4.2  & 39.970688 & -1.5763221 & 1.2728   \\
4.3  & 39.961971 & -1.5779671 & 1.2728   \\
\hline
5.1  & 39.973473 & -1.5890463 & 1.2775   \\
5.2  & 39.970576 & -1.5891946 & 1.2775   \\
5.3  & 39.969472 & -1.5890961 & 1.2775   \\
5.4  & 39.96858  & -1.5890045 & 1.2775   \\
\hline
6.1  & 39.979641 & -1.5770904 & 1.0633   \\
6.2  & 39.969405 & -1.5771811 & 1.0633   \\
6.3  & 39.964334 & -1.5782307 & 1.0633   \\
\hline
7.1  & 39.986567 & -1.5775688 & 2.7512   \\
7.2  & 39.969882 & -1.5807608 & 2.7512   \\
7.3  & 39.969788 & -1.5804299 & 2.7512   \\
7.4  & 39.969046 & -1.5774833 & 2.7512   \\
7.5  & 39.968815 & -1.5856313 & 2.7512   \\
7.6  & 39.968454 & -1.5715778 & 2.7512   \\
7.7  & 39.968142 & -1.5710778 & 2.7512   \\
7.8  & 39.961533 & -1.5800028 & 2.7512   \\
\hline
8.1  & 39.964485 & -1.5698065 & 2.884*    \\
8.2  & 39.961889 & -1.5736473 & 2.884*    \\
\hline
9.1  & 39.982022 & -1.5765337 & 1.5182   \\
9.2  & 39.969486 & -1.5762654 & 1.5182   \\
9.3  & 39.962402 & -1.5778911 & 1.5182   \\
\hline
10.1 & 39.98846  & -1.5719676 & 7.04*    \\
10.2 & 39.963839 & -1.5693802 & 7.04*    \\
10.3 & 39.960789 & -1.5741702 & 7.04*   \\
\hline
11.1 & 39.9841   & -1.5709127 & 3.4809   \\
11.2 & 39.969682 & -1.566636  & 3.4809   \\
11.3 & 39.959198 & -1.5753221 & 3.4809   \\
\hline
12.1 & 39.979513 & -1.5717782 & 4.248    \\
12.2 & 39.97521  & -1.5688203 & 4.248    \\
12.3 & 39.956759 & -1.5775032 & 4.248    \\
\hline
13.1 & 39.981313 & -1.5782202 & 3.1309   \\
13.2 & 39.974254 & -1.585577  & 3.1309   \\
13.3 & 39.972309 & -1.578091  & 3.1309   \\
13.4 & 39.972192 & -1.5801027 & 3.1309   \\
13.5 & 39.957673 & -1.580459  & 3.1309   \\
\hline
14.1 & 39.984008 & -1.5784556 & 3.7084   \\
14.2 & 39.971935 & -1.5870512 & 3.7084   \\
14.3 & 39.971328 & -1.580604  & 3.7084   \\
14.4 & 39.971027 & -1.5777907 & 3.7084   \\
14.5 & 39.970450 & -1.5689437 & 3.7084   \\
14.6 & 39.970391 & -1.5696387 & 3.7084   \\
14.7 & 39.958795 & -1.5805488 & 3.7084   \\
\hline
15.1 & 39.984414 & -1.5841111 & 3.7743   \\
15.2 & 39.964016 & -1.5880782 & 3.7743   \\
\hline
16.1 & 39.985403 & -1.5808406 & 4.2567   \\
16.2 & 39.969758 & -1.5885333 & 4.2567   \\
16.3 & 39.960235 & -1.5836508 & 4.2567   \\
\hline
17.1 & 39.981476 & -1.5820728 & 4.4296   \\
17.2 & 39.97583  & -1.5870613 & 4.4296   \\
17.3 & 39.957362 & -1.5820861 & 4.4296   \\
\hline
18.1 & 39.985142 & -1.5790944 & 5.6493   \\
18.2 & 39.971996 & -1.5878654 & 5.6493   \\
18.3 & 39.958316 & -1.5813093 & 5.6493   \\
\hline
19.1 & 39.965279 & -1.5878055 & 5.7505   \\
19.2 & 39.963619 & -1.5868798 & 5.7505   \\
\hline
20.1 & 39.981539 & -1.5814028 & 1.2567   \\
20.2 & 39.967252 & -1.5849694 & 1.2567   \\
20.3 & 39.966733 & -1.5846943 & 1.2567   \\
\hline
21.1 & 39.981675 & -1.5796852 & 3.1309   \\
21.2 & 39.974406 & -1.5861017 & 3.1309   \\
21.3 & 39.957906 & -1.5810108 & 3.1309   \\
\hline
22.1 & 39.980113 & -1.5667264 & 5.9386   \\
22.2 & 39.977166 & -1.5662748 & 5.9386   \\
22.3 & 39.957315 & -1.572744  & 5.9386   \\
\hline
23.1 & 39.963111 & -1.570603  & 4.916    \\
23.2 & 39.96204  & -1.5723407 & 4.916    \\
\hline
24.1 & 39.987084 & -1.5790992 & 3.8145   \\
24.2 & 39.966982 & -1.5867999 & 3.8145   \\
24.3 & 39.961703 & -1.5832126 & 3.8145   \\
\hline
25.1 & 39.979924 & -1.571393  & 3.9359   \\
25.2 & 39.97446  & -1.5680963 & 3.9359   \\
25.3 & 39.95717  & -1.5769717 & 3.9359   \\
\hline
26.1 & 39.980691 & -1.5711198 & 3.0161   \\
26.2 & 39.97244  & -1.5671511 & 3.0161   \\
26.3 & 39.958292 & -1.5759068 & 3.0161   \\
\hline
27.1 & 39.987817 & -1.5774528 & 2.9101   \\
27.2 & 39.967058 & -1.5845583 & 2.9101   \\
27.3 & 39.963492 & -1.5822806 & 2.9101   \\
\hline
28.1 & 39.983596 & -1.5674774 & 4.4897   \\
28.2 & 39.968425 & -1.5646657 & 4.4897   \\
28.3 & 39.960838 & -1.5691328 & 4.4897   \\
\hline
29.1 & 39.983351 & -1.5704081 & 5.6459   \\
29.2 & 39.972404 & -1.5663533 & 5.6459   \\
\hline
30.1 & 39.980667 & -1.5747346 & 5.4476   \\
30.2 & 39.97475  & -1.5693301 & 5.4476   \\
30.3 & 39.956156 & -1.5786786 & 5.4476   \\
\hline
31.1 & 39.988097 & -1.5751871 & 4.4953   \\
31.2 & 39.966285 & -1.5693446 & 4.4953   \\
31.3 & 39.960682 & -1.5783795 & 4.4953   \\
\hline
32.1 & 39.966215 & -1.5879961 & 4.882    \\
32.2 & 39.962722 & -1.5860036 & 4.882    \\
\hline
33.1 & 39.985048 & -1.579559  & 5.2437   \\
33.2 & 39.971805 & -1.5880395 & 5.2437   \\
33.3 & 39.970107 & -1.5701499 & 5.2437   \\
33.4 & 39.958566 & -1.5817008 & 5.2437   \\
\hline
34.1 & 39.981538 & -1.5658624 & 6.1735   \\
34.2 & 39.975825 & -1.5644423 & 6.1735   \\
\hline
35.1 & 39.962444 & -1.5807098 & 6.2855   \\
35.2 & 39.965996 & -1.5843845 & 6.2855   \\
\hline
36.1 & 39.970391 & -1.5687943 & 5.6489   \\
36.2 & 39.970428 & -1.5694203 & 5.6489   \\
\hline
37.1 & 39.977154 & -1.5737917 & 3.2      \\
37.2 & 39.975063 & -1.5721045 & 3.2      \\
\hline
38.1 & 39.983162 & -1.5796664 & 4.3381   \\
38.2 & 39.973383 & -1.5874465 & 4.3381   \\
38.3 & 39.970632 & -1.5710393 & 4.3381   \\
38.4 & 39.957967 & -1.5815081  & 4.3381   \\
\hline
39.1 & 39.982296 & -1.576975  & 1.272*   \\
39.2 & 39.967933 & -1.5773472 & 1.272*   \\
39.3 & 39.965442 & -1.5780222 & 1.272*   \\
\hline
40.1 & 39.963579 & -1.5656333 & 1.0315   \\
40.2 & 39.963375 & -1.5659528 & 1.0315   \\
40.3 & 39.962958 & -1.5661111 & 1.0315   \\
\hline
41.1 & 39.984054 & -1.5733556 & 1.973*   \\
41.2 & 39.966563 & -1.5696694 & 1.973*   \\
41.3 & 39.962117 & -1.57525   & 1.973*   \\
\hline
42.1 & 39.977717 & -1.5827917 & 2.336*   \\
42.2 & 39.976646 & -1.5836028 & 2.336*   \\
\hline
43.1 & 39.982400 & -1.5811    & 8.593*   \\
43.2 & 39.975875 & -1.58726   & 8.593*  \\
\hline
\enddata
\tablenotetext{1}{Refer to Table~\ref{tab: knot_counterpart} for the positions of Systems 2.2--2.10.}
\label{tab: System_list}
\end{deluxetable}

\section{Spectra of knot counterparts in the Dragon arc}
\label{sec: Dragon_Arc_spectra}
\begin{figure*}[h]
    \centering
    \includegraphics[width = \textwidth]{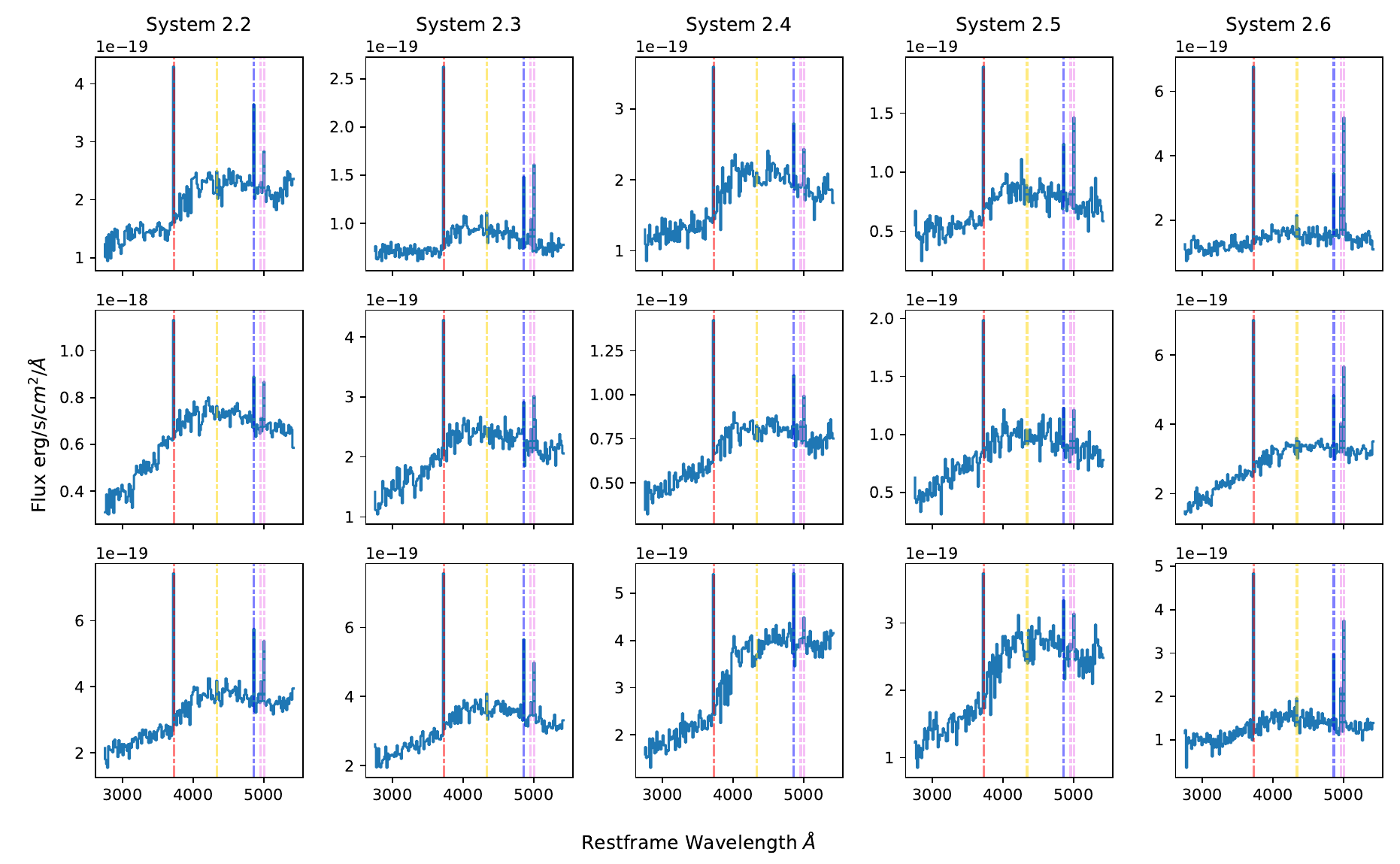}
    \caption{Rest-frame MUSE spectra of all the bright knots identified as part of System 2.  Prominent emission lines are indicated by vertical dashed lines, for which red corresponds to the [O~II] doublet, yellow corresponds to H$\gamma$, blue corresponds to H$\beta$, and pink corresponds to the [O~III] doublet.}
    \label{fig: spectra_all}
\end{figure*}

Figure\,\ref{fig: spectra_all} shows spectra, from the MUSE-VLT, towards each lensed counterpart of systems 2.2--2.6.  Each system comprises a sets of knots in the Dragon arc as observed with the HST.  Notice that all the spectra display emission lines characteristic of HII regions.  Because the MUSE-VLT spectra were taken at a much lower angular resolution than the HST images, we cannot be sure that an individual lensed counterpart is truly associated with a line-emitting region.  Nonetheless, that every lensed counterpart is associated with a line-emitting region provides reassurance that each system corresponds to young star cluster still displaying a HII region.

\section{Supplementary Materials of Microlensing Simulation}

We now present additional information related to our microlensing simulation, and discuss certain aspects of the simulation in more detail.  Most importantly, we describe the tests carried out to validate the assumptions made in our simulation.  We end by discussing some of the more important caveats, as well as limitations, of our simulations.

\subsection{Stellar Surface Mass Density at Knot 2.6.2}\label{sec: muL_CM}

The cluster member enclosed by the black circle labelled 5 in the lower panel of Fig.\,\ref{fig: RGB} lies in the close vicinity of knot 2.6.2, which is indicated by a green arrow that partially lies within the same black circle.   This cluster member contributes significantly to the surface mass density of stars at the position of knot 2.6.2.  To estimate its stellar contribution, we first fit a model stellar population having a Kroupa IMF to its SED using {\tt Bagpipes} \citep{Carnall_2018}.  For an exponentially decaying star-formation rate over time, which presents a marginally better fit to the SED than the other star-formation histories that we tried, the total mass of stars in this galaxy that have survived to the present epoch is $M_{\rm stars} = 3.47 \times 10^{9}$\,M$_{\odot}$.

To estimate the stellar surface mass density of this cluster member at the position of knot 2.6.2, we fit a 2-D S\'ersic profile to its light in the F160W image.  This image was chosen as it provides the highest contrast between the cluster member of interest and the Dragon arc.  By adopting a constant mass-to-light ratio and based on the total stellar mass derived in the manner described above, we then converted the radial light profile of this cluster member to a radial mass profile.  In this way, we find a contribution from this cluster member to the stellar surface mass density at knot 2.6.2 of $\sim$40$\,M_{\odot}\,{\rm pc}^{-2}$ (corresponding to $= 0.011 \kappa_{\star}$).



\subsection{Simulation Field-of-View}

As mentioned in Section~\ref{sec: simulation}, the field of view (FOV) employed in our simulation of $0.8\ \textrm{pc} \times 0.8\ \textrm{pc}$ (as measured in the source plane) is likely to be much smaller than the sizes of star clusters in the Dragon arc.  (The multiply-lensed knots of interest, corresponding to star clusters in the Dragon arc, are mostly unresolved, for which we place upper limits on their sizes of $\sim$$200\,\textrm{pc}$ given the PSF of the HST and lensing magnifications involved).
Within the simulation FOV, stars belonging to a star cluster is macrolensed by A370 (imposing a constant magnification across the FOV) and microlensed by stars (and, when incorporated, PBHs) in the galaxy cluster -- for which the simulation FOV is sufficiently large to contain a representative mass distribution of stellar microlenses (as these stars are relatively old and therefore span only a relatively narrow mass range).  In our simulation, we effectively use the same representation of stellar microlenses (over the $0.8\ \textrm{pc} \times 0.8\ \textrm{pc}$ FOV) multiple times to span the region occupied by the star cluster.  Equivalently, to save computation time, what we actually do is to distribute (at random) stars having a Kroupa IMF over the computational FOV until reaching the total mass as inferred for star clusters of interest in the Dragon arc (see more details in Section~\ref{sec: simulation}).

\begin{figure}[ht!]
    \centering
    \includegraphics[width = 0.5\textwidth]{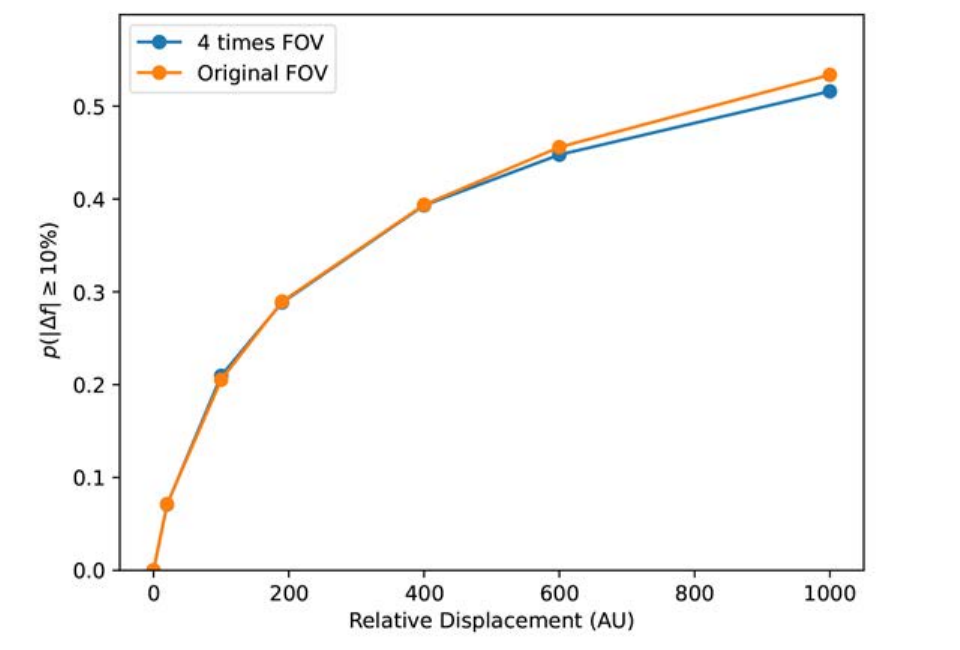}
    \caption{$p(|\Delta f| \ge 10\%)$ as a function of relative displacement between microlenses and lensed star cluster for two different simulation FOVs.  Blue symbols are for a FOV that is twice the dimensions and hence four times the area of the FOV represented by the orange symbols.   Both simulations are for $\mu = -13.5$ and $\kappa_{\star} = 0.014$ (and no PBHs). The results are closely identical except at large relative displacements -- much larger than the displacement corresponding to the Flashlights observations of A370 -- whereby a larger fraction of lensed stars are displaced outside the smaller simulation FOV compared with the larger simulation FOV.}
    \label{fig: FOV_test}
\end{figure}

To test whether using a larger simulation FOV would give a similar result, we conducted one simulation (using the parameters $\kappa_{\star} = 0.014$, $\mu = -13.5$, and $f(M) = 0$) for which the simulation FOV employed is $1.6\,\textrm{pc} \times 1.6\,\textrm{pc}$.  The results are shown in Figure~\ref{fig: FOV_test}, for which the orange curve is for a simulation FOV of $0.8\ \textrm{pc} \times 0.8\ \textrm{pc}$ and the blue curve a larger simulation FOV of $1.6\,\textrm{pc} \times 1.6\,\textrm{pc}$.  As is apparent, the results are closely similar, showing appreciable differences only at large displacements between two observational epochs that are much larger than the displacement between the two epochs of our Flashlights observations of A370.  At such large displacements, a larger fraction of lensed stars are displaced outside the smaller simulation FOV compared with the larger simulation FOV, explaining the differences between the two simulation FOVs at large relative displacements.

\subsection{Semianalytical Approximation}

\subsubsection{Dependence on Effective Surface Mass Density}
\label{sec: kappa_effect}

Figure\,\ref{fig: combined_motion} shows that both $\sigma_{L}$ and $p(|\Delta f| \ge 10\%)$ increases with the effective surface mass density $\Sigma_{\textrm{eff}} = \mu \Sigma_*$ (for stellar microlenses only) or $\Sigma_{\textrm{eff}} = \mu (\Sigma_* + \Sigma_{\rm PBH})$ (for both stellar and PBH microlenses).  This trend is as expected because an increase in $\Sigma_{\textrm{eff}}$ results in a denser network of microcaustics, and consequently also a greater likelihood of either magnification or demagnification of individual background stars (belonging to a star cluster) across this network.  Both $\sigma_{L}$ and $p(|\Delta f| \ge 10\%)$ also increases with a larger displacement between the lensed star cluster and the microlenses owing to a longer time interval between observations.  This trend also is as expected because a larger displacement results in a greater likelihood of background stars experiencing a larger change in magnification, and consequently for the star cluster as a whole to display a larger variation in total brightness.


The aforementioned trend, however, cannot continue to indefinitely large $\Sigma_{\textrm{eff}}$.  As $\Sigma_{\textrm{eff}}$ increases, the microcaustics become so densely distributed that they eventually overlap, reducing the likelihood of background stars experiencing a significant change in magnification between two observation epochs.  As a consequence, both $\sigma_{L}$ and $p(|\Delta f| \ge 10\%)$ is expected to reach a maximum at large $\Sigma_{\textrm{eff}}$ and perhaps even decrease thereafter.  Previous studies \citep{Oguri_2018, Diego_2018, Palencia_23} have identified $\Sigma_{\textrm{eff}} = \Sigma_{\textrm{crit}}$ ($\kappa_{\textrm{eff}} = 1$) as to the regime at which microcaustics begin to overlap.  At $\Sigma_{\textrm{eff}} > \Sigma_{\textrm{crit}}$, stars sweeping through the network of microcaustics experience smaller changes in their individual magnifications, resulting in decreased levels of brightness variations for the stellar ensemble (star cluster) as a whole.  As demonstrated in Figure~\ref{fig: pvD} (right panel), at a fixed $\delta$ (related to the time interval between observations), $p(|\Delta f| \ge 10\%)$ increases rapidly for small $\kappa_{\textrm{eff}}$ but gradually levels off as $\kappa_{\textrm{eff}}$ approaches unity.  The latter behavior had been noticed in earlier work, where $\kappa_{\textrm{eff}}\approx 1$ corresponds to the saturation regime described by \cite{Diego_2018} at which an increase in either $\mu$ or $\Sigma_{*}$ has little impact on the statistics of magnification.

\subsubsection{Effect of Relative Displacement}

Figure\,\ref{fig: pvD} (left panel) shows that, at a fixed $\kappa_{\textrm{eff}}$, $p(|\Delta f| \ge 10\%)$ increases rapidly with $\delta$ before levelling off.  Initially, a larger relative displacement causes lensed stars near microcaustics to undergo a larger change in magnification, and therefore for the stellar ensemble to more likely display an overall change in brightness.  When this relative displacement becomes too large, however, the same lensed stars move away entirely from the microcaustics to regions of low magnifications.  As the latter regions occupy a much larger area in total than regions of high magnifications (see Figure~\ref{fig: microcaustics}), even larger relative displacements make no difference to changes in the magnifications of lensed stars originally near microcaustics -- causing $p(|\Delta f| \ge 10\%)$ to level of at sufficiently large $\delta$.


Figure\,\ref{fig: pvD} (left panel) also shows that, as $\kappa_{\textrm{eff}}$ increases, $p(|\Delta f| \ge 10\%)$ levels off at smaller values of $\delta$.  As $\kappa_{\textrm{eff}}$ increases, microcaustics become more closely packed, so that regions away from microcaustics much more quickly reach a nearly constant magnification.  As a consequence, individual stars originally located near microcaustics experience no further change in their individual magnifications at a smaller relative displacement between microlenses and lensed stars when $\kappa_{\textrm{eff}}$ is larger, resulting in $p(|\Delta f| \ge 10\%)$ leveling off at smaller values of $\delta$.

\subsection{Stellar Evolution and the Abundance of Hyperluminous Stars}

As mentioned in Section~\ref{sec: IMF}, our simulated lensed star cluster has a mass function described by Kroupa IMF truncated at stellar mass of $100 \rm \, M_\sun$.  Such a model star cluster provides a reasonable description of star clusters that still exhibit HII regions and are therefore relatively young, such as those considered in the Dragon arc (estimated ages of $\sim$$4\, \textrm{Myr}$).  Of course, real star clusters may have mass functions that deviate from a Kroupa IMF and/or be truncated at a different stellar mass -- and may well have evolved stars more luminous than main-sequence stars, especially as the star cluster ages.  As mentioned in Section~\ref{sec: simulation}, luminous stars have an especially large effect on the likelihood of brightness changes in the lensed star cluster when subjected to microlensing.



\begin{figure}[ht!]
    \centering
    \includegraphics[width = \linewidth]{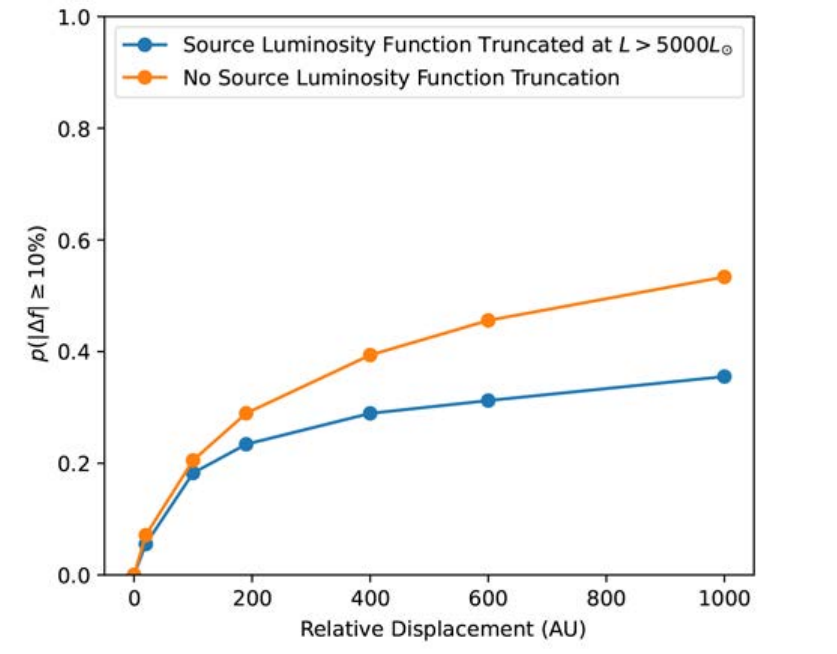}
    \caption{$p(|\Delta f| \ge 10\%)$ as a function of relative displacement between microlenses and lensed star cluster for $\mu = -13.5$ and $\kappa_{\star} = 0.014$.  Blue symbols are for a lensed star cluster truncated at a stellar mass of $20\, {\rm M}_{\odot}$ and hence a stellar luminosity of $5000\, {\rm L}_{\odot}$.  Orange symbols are for a lensed star cluster truncated at a stellar mass of $100\, {\rm M}_{\odot}$ and hence a stellar luminosity of $3.2 \times 10^{6}\, {\rm L}_{\odot}$.}
    \label{fig: source_trun_motion}
\end{figure}

As a starting point to examine how our results depend on the assumed physical properties of lensed star clusters, we now truncate the stellar mass of the lensed star cluster at $20\, {\rm M}_{\odot}$ (for which stellar libraries allow for no evolved hyperluminous stars), but which otherwise have the same physical properties as before.  Figure~\ref{fig: source_trun_motion} shows how $p(|\Delta f| \ge 10\%)$ changes with relative displacement between the microlenses and lensed star cluster for $\mu = -13.5$ and $\kappa_{\star} = 0.014$, whereby blue symbols are for the lensed star cluster truncated at $20\, {\rm M}_{\odot}$ and orange symbols truncated at $100\, {\rm M}_{\odot}$.  As can be seen, truncation at a lower stellar mass results in a lower $p(|\Delta f| \ge 10\%)$ as would be expected given the lower maximal luminosity of stars in the lensed star cluster.  More interestingly, once the relative displacement exceeds about 100\,AU, the change in $p(|\Delta f| \ge 10\%)$ with relative displacement becomes much slower for the lensed star cluster truncated at $20\, {\rm M}_{\odot}$ compared with that truncated at $100\, {\rm M}_{\odot}$.  We therefore anticipate observations over longer time scales to have greater sensitivity to the presence of luminous stars in lensed star clusters subjected to microlensing.

Of course, the situation becomes more complicated once we allow for stellar evolution and hence the age of the star cluster under consideration (in addition to the stellar IMF and the truncation in stellar mass) -- especially when hyperluminous stars are present.  Nonetheless, we now have the tools and have introduced the parameters for quantifying the probability of a given brightness change in a lensed star cluster subjected to microlensing.  The same work as described in this paper can be repeated for star clusters having a different stellar IMF, truncation at a different stellar mass, and age given a stellar evolutionary library. 

\end{document}